\documentclass[usenatbib]{mn2e}
\usepackage{lscape}
\usepackage{amsmath,amssymb}
\usepackage{color}
\usepackage{graphicx}
\usepackage{verbatim}

\usepackage{txfonts}
\usepackage[T1]{fontenc}
\usepackage{aecompl}

\usepackage[dvipsnames]{xcolor}

\usepackage[bookmarks=true,pdfhighlight=/I]{hyperref}
\hypersetup{
  colorlinks=true,
  pdfborder=2 2 0.,
  linkcolor=BrickRed,
  anchorcolor=black,
  citecolor=Blue,
  urlcolor=black,
}

\catcode`\@=11
\def\gta{\ifmmode{\,\mathrel{\mathpalette\@versim>\,}}
    \else{$\,\mathrel{\mathpalette\@versim>}\,$}\fi}
\def\lta{\ifmmode{\,\mathrel{\mathpalette\@versim<\,}}
    \else{$\,\mathrel{\mathpalette\@versim<}\,$}\fi}
\def\@versim#1#2{\lower 2.9truept \vbox{\baselineskip 0pt \lineskip
    0.5truept \ialign{$\m@th#1\hfil##\hfil$\crcr#2\crcr\sim\crcr}}}
\catcode`\@=12  

\makeatletter
\def\blfootnote{\xdef\@thefnmark{}\@footnotetext}
\makeatother

\newcommand{\Msol}{M_\odot}
\newcommand{\Mpart}{M_{\rm{part}}}
\newcommand{\Hubble}{H_0}

\newcommand{\Omegazerom}{\Omega_{0, \rm m}}
\newcommand{\Omegazerolambda}{\Omega_{0, \Lambda}}

\newcommand{\rhoc}{\rho_{\rm c}}
\newcommand{\Deltac}{\Delta_{\rm c}}
\newcommand{\barv}{\bar{v}}
\newcommand{\sigmav}{\sigma_{\rm V}}
\newcommand{\sigmazero}{\sigma_0}

\newcommand{\Eorb}{E_{\rm orb}}
\newcommand{\rg}{r_{\rm g}}
\newcommand{\rh}{r_{\rm h}}
\newcommand{\re}{R_{\rm e}}

\newcommand{\rdelta}{r_\Delta}
\newcommand{\sigmatwelve}{\sigma_{\rm V,12}}
\newcommand{\rhtwelve}{r_{\rm h,12}}
\newcommand{\EPMA}{E_{\rm orb,PMA}}
\newcommand{\UPMA}{U_{\rm orb,PMA}}

\newcommand{\Uorb}{U_{\rm orb}}
\newcommand{\rhf}{r_{\rm h,f}}
\newcommand{\rhi}{r_{\rm h,1}}
\newcommand{\sigmavf}{\sigma_{\rm V,f}}
\newcommand{\sigmavi}{\sigma_{\rm V,1}}

\newcommand{\drel}{d_{\rm rel}}

\newcommand{\Lorb}{L_{\rm orb}}
\newcommand{\AHF}{{\sc AHF }}
\newcommand{\Gadget}{{\sc Gadget-2 }}

\newcommand{\Mstar}{M_\ast}

\newcommand{\sigmastar}{\sigma_\ast}
\newcommand{\rconv}{r_{\rm conv}}
\newcommand{\Mlower}{M_{\rm lower}}

\defcitealias{Behroozi+2010}{B10}
\defcitealias{Leauthaud+2012}{L12}
\defcitealias{Cimatti+2012}{CNC}
\defcitealias{Nipoti+2012}{N12}
\defcitealias{Mo+1998}{Mo et al. 1998}
\defcitealias{Nipoti+2003}{Nipoti et al. 2003}
\defcitealias{Hilz+2013}{Hilz et al. 2013}
\defcitealias{Belli+2013}{Belli et al. 2013}

\begin{document}
\date{Accepted 2014 February 12. Received 2014 February 12; in original form 2013 October 4.}
\title[Halo-galaxy evolution]
{The imprint of dark matter haloes on the size and velocity dispersion evolution of early-type galaxies}
\author[L. Posti, C. Nipoti, M. Stiavelli and L. Ciotti]{Lorenzo Posti$^{1,2}$\thanks{E-mail: lorenzo.posti@unibo.it},
Carlo Nipoti$^{1}$, Massimo Stiavelli$^{2}$ and Luca Ciotti$^{1}$
\\ \\
$^{1}$Dipartimento di Fisica e Astronomia, Universit\`a di Bologna, viale Berti-Pichat 6/2, I-40127 Bologna, Italy\\
$^{2}$Space Telescope Science Institute, 3700 San Martin Drive, Baltimore, MD 21218, USA}

\maketitle
\begin{abstract}
Early-type galaxies (ETGs) are observed to be more compact, on
average, at $z \gtrsim 2$ than at $z\simeq0$, at fixed stellar
mass. Recent observational works suggest that such size evolution
could reflect the similar evolution of the host dark matter halo density
as a function of the time of galaxy quenching. We explore this
hypothesis by studying the distribution of halo central
velocity dispersion ($\sigmazero$) and half-mass radius ($\rh$) as
functions of halo mass $M$ and redshift $z$, in a cosmological
$\Lambda$-CDM $N$-body simulation. In the range
$0\lesssim z\lesssim 2.5$, we find $\sigmazero\propto M^{0.31-0.37}$ and
$\rh\propto M^{0.28-0.32}$, close to the values expected for homologous
virialized systems. At fixed $M$ in the range $10^{11}\Msol \lesssim M\lesssim
5.5 \times 10^{14}\Msol$ we find $\sigmazero\propto(1+z)^{0.35}$ and
$\rh\propto(1+z)^{-0.7}$. We show that such evolution of the halo scaling
laws is driven by individual haloes growing in mass following the
evolutionary tracks $\sigmazero\propto M^{0.2}$ and $\rh\propto M^{0.6}$,
consistent with simple dissipationless merging models in which the
encounter orbital energy is accounted for. We compare the $N$-body
data with ETGs observed at $0\lesssim z\lesssim3$ by populating the
haloes with a stellar component under simple but justified assumptions:
the resulting galaxies evolve consistently with the
observed ETGs up to $z \simeq 2$, but the model has
difficulty reproducing the fast evolution observed at $z\gtrsim2$.
We conclude that a substantial fraction
of the size evolution of ETGs can be ascribed to a systematic
dependence on redshift of the dark matter haloes structural properties.
\end{abstract}
\begin{keywords}
galaxies: haloes - galaxies: formation - galaxies: evolution - cosmology: dark matter
\end{keywords}

\section{Introduction}
\label{sec:intro}
Since early studies in the '$70$s, we know that early-type galaxies
(ETGs) adhere to some empirical scaling relations, such as the
luminosity-velocity dispersion \citep{FaberJackson1976}, size-surface
brightness \citep{Kormendy1977}, Fundamental Plane
\citep[][]{DjorgovskiDavis1987,Dressler+1987}, black-hole mass-bulge
mass \citep{Magorrian+1998}, black-hole mass-velocity dispersion
\citep[][]{FerrareseMerritt2000,Gebhardt+2000} and
black-hole mass-S\'ersic index
\citep[][]{Graham+2001,GrahamDriver2007} relations.  Such
correlations, some of which were first used as distance estimators to
help building the distance scale ladder, have given the astrophysical
community important clues about the possible scenarios of galaxy
formation \citep[e.g.,][]{Ciotti2009}.  In this respect, the stellar
mass-size relation has currently a special role, because galaxy sizes
and masses can be measured out to $z \simeq 2.5-3$. With such data
available, different authors found indications that the population
of ETGs undergoes a significant size evolution from $z\simeq 3 $ to
$z \simeq 0$, such that present-day galaxies have, on average, significantly
larger sizes than higher $z$ galaxies of similar stellar mass \citep[see
e.g.,][]{Stiavelli+1999,Ferguson+2004,Daddi+2005,Trujillo+2006,
Cimatti+2008,vanderWel+2008,vanDokkum+2008,Saracco+2009,Cassata+2011,
Damjanov+2011,Krogager+2013}.
For the current galaxy formation models it is still challenging and
non-trivial to explain such behaviour of massive ETGs. Various
mechanisms have been proposed to explain the observed size evolution,
including dry (i.e., dissipationless) major and minor merging
\citep[see][]
{KhochfarSilk2006,Nipoti+2009b,Naab+2009,Hopkins+2009b,
Lopez-Sanjuan+2012,Laporte+2013,Sonnenfeld+2013}
and feedback-driven expansion
\citep[see e.g.,][]{Fan+2008,Fan+2010,Ragone-Figueroa+2011,Ishibashi+2013}.
Currently, the issue is far from being resolved and further
observations, together with more comprehensive theoretical models, are
desirable.

Recently, \citet[][see also \citeauthor{Poggianti+2013}
\citeyear{Poggianti+2013}]{Carollo+2013} argued that the median size
growth of ETGs of stellar mass $10^{10.5}\Msol \leq \Mstar \leq 10^{11}
\Msol$ could be due to the dilution of the sample of galaxies quenched
at early times in a population of bluer and larger galaxies
that have been quenched much later.
In the sample of ETGs with $\Mstar>10^{11}\Msol$ the same authors find
indications of intrinsic size evolution, which can not be explained
with the dilution of the population.
In other words, \cite{Carollo+2013} find evidence
that not all the progenitors of local quenched-ETGs can be
identified with the compact quiescent ETGs observed at $z\simeq 1-2$,
because a substantial fraction of present-day ETGs have stopped
forming stars much later than the higher-$z$ ETGs. An interesting
conclusion of \cite{Carollo+2013} is that the stellar
density of ETGs scales with the mean density of the Universe
\emph{at the time of quenching}.
This suggests that the host halo evolution could be the main driver
of the galaxy evolution, in the sense that the redshift-dependence of
the properties of an ETG results similar to that of its host halo
\citep[see also recent results by][]{Stringer+2013}.

The natural tool to explore such halo-galaxy connection would be
a large-scale, high-resolution, cosmological simulation jointly following
the evolution of dark matter (hereafter DM) and baryons, including star
formation and feedback. However, given the well known uncertainties and
technical issues still present in this method \citep[see
e.g.,][]{Keres+2012,Vogelsberger+2012,Hopkins+2013}, we adopt here
a simpler approach, trying to extract useful information on the
evolution of ETGs studying the behaviour of a population of DM
haloes in a DM-only cosmological simulation. We focus our attention
on the scaling laws of DM haloes in a $\Lambda$-CDM Universe and in
particular on their size and velocity dispersion evolution.
Our aim is trying to understand whether the evolution of the haloes is
somehow similar to that of ETGs which are expected to be hosted in
such haloes. Therefore, we will also try to compare our $N$-body
data with available observations of ETGs, populating haloes with
galaxies under simple but justified assumptions.

This paper is organized as follows: in Section \ref{sec:methods_defs}
we present the methods of our investigation and we set the stage with
all the definitions and simple theoretical expectations; in Section
\ref{sec:tot_scaling} we show our results on the scaling relations of
the dark-halo population; in Section \ref{sec:evo_sing} we trace the
merger histories of individual haloes which are representative for the
entire population; in Section \ref{sec:DMdriven_evo} we link the
dark-halo properties with those of the ETGs and compare the predicted
size and velocity dispersion evolution with recent observations;
Section \ref{sec:concl} summarizes and concludes.

\section{Methods and definitions}
\label{sec:methods_defs}
\subsection{Computational tools}
\label{sec:comp_tools}
We performed a cosmological $N$-body simulation with the
publicly available code \Gadget
\citep[][see also \citeauthor{Springel+2001} \citeyear{Springel+2001}]
{Springel2005} in a standard
$\Lambda$-CDM flat Universe where the matter density, dark-energy
density and Hubble constant are, respectively, $\Omegazerom=0.28$,
$\Omegazerolambda=0.72$ and $\Hubble=70\,{\rm km\, s^{-1} \,
Mpc^{-1}}$. In the run we simulated the evolution of
$512^3 \simeq 1.3\times 10^8$ particles of mass
$\Mpart\simeq 1.5\times 10^9\Msol/h$,
from $z=99$ to $z=0$, in a cosmological comoving box of side
$l=128\,{\rm Mpc}/h$, where $h$ is the reduced Hubble constant
$h=\Hubble / 100 \,\rm{km\, s^{-1}\, Mpc^{-1}}$.
The initial conditions of the simulation were generated
using a modified version of the publicly available code {\sc Grafic2}
\citep[see][]{Bertschinger2001}.  We used a softening length of $\sim
1\,{\rm kpc}/h$ throughout the simulation.
The simulation was run on $72$ cores on the UDF Linux cluster at STScI
(Baltimore) and took about $6$ days to complete.  We produced $16$
snapshots equally spaced in $\log a$, where $a(t)=(1+z)^{-1}$ is the
cosmic scale factor, from $z \simeq 2.5$ to $z=0$.

\begin{figure}
\includegraphics[width=\hsize]{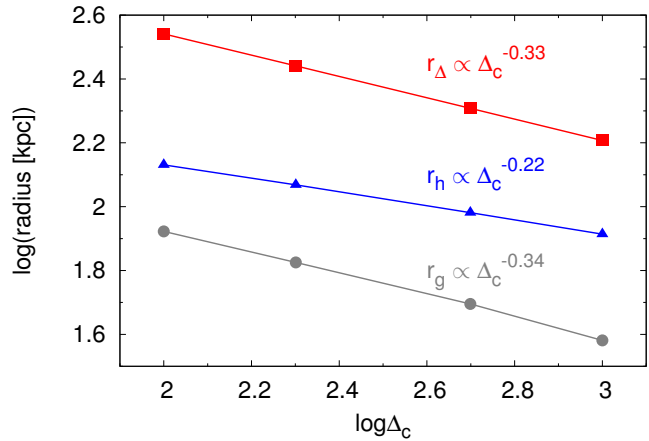}
\caption{Virial radius 
  (red squares, see equation \ref{def:halo}), gravitational radius
  (gray circles, see equation \ref{def:rg}) and half-mass radius
  (blue triangles) as functions of the critical overdensity $\Deltac$
  for a $M\simeq 10^{13}\Msol$ halo, taken from the $z=0$ snapshot of
  the simulation.}
\label{fig:size-delta}
\end{figure}

\begin{figure*}
\includegraphics[width=0.33\hsize]{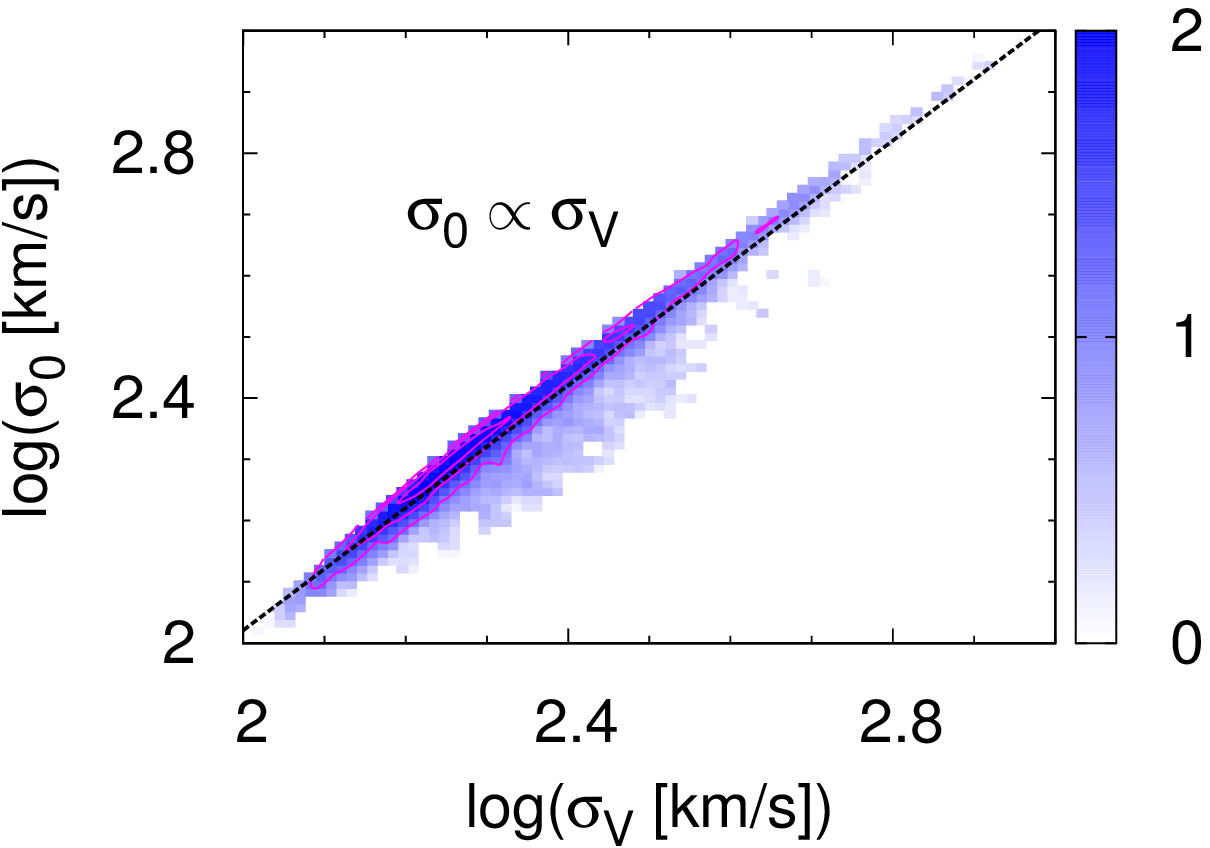}
\includegraphics[width=0.33\hsize]{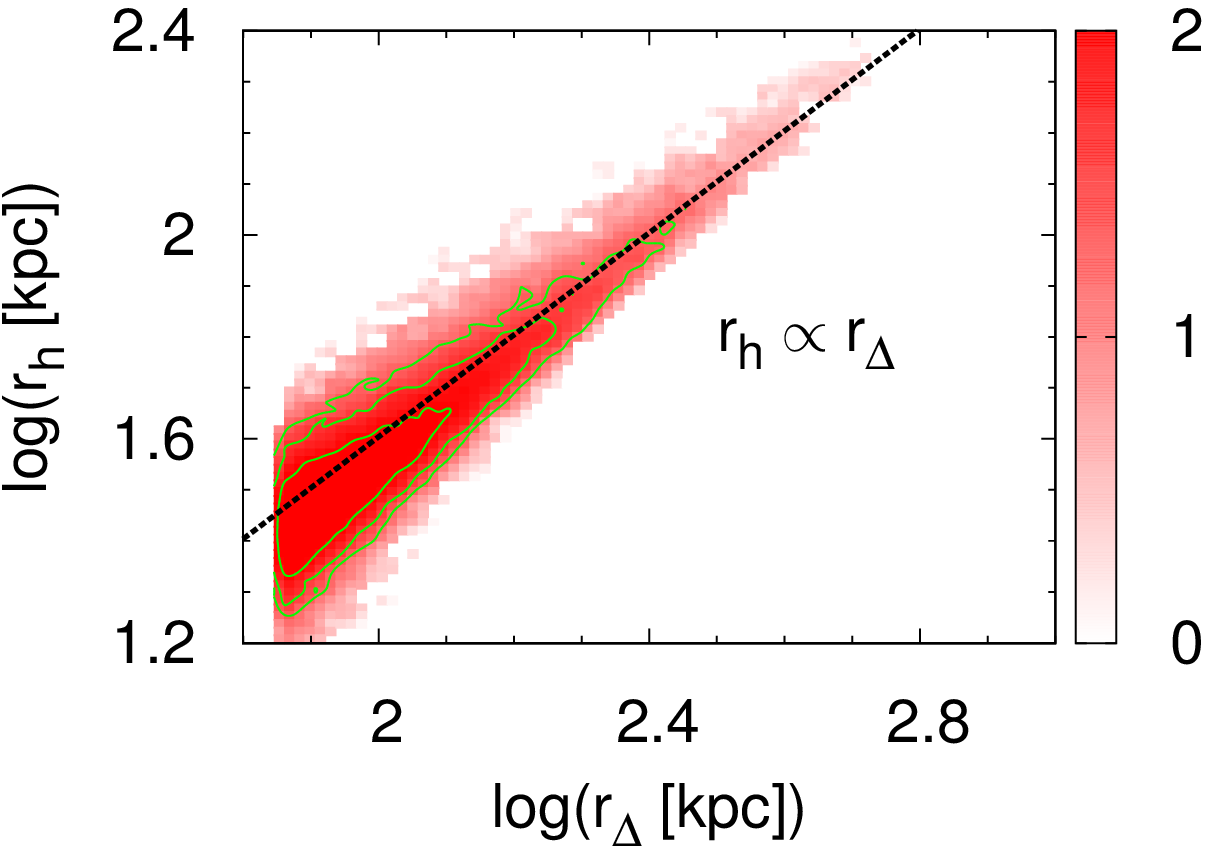}
\includegraphics[width=0.33\hsize]{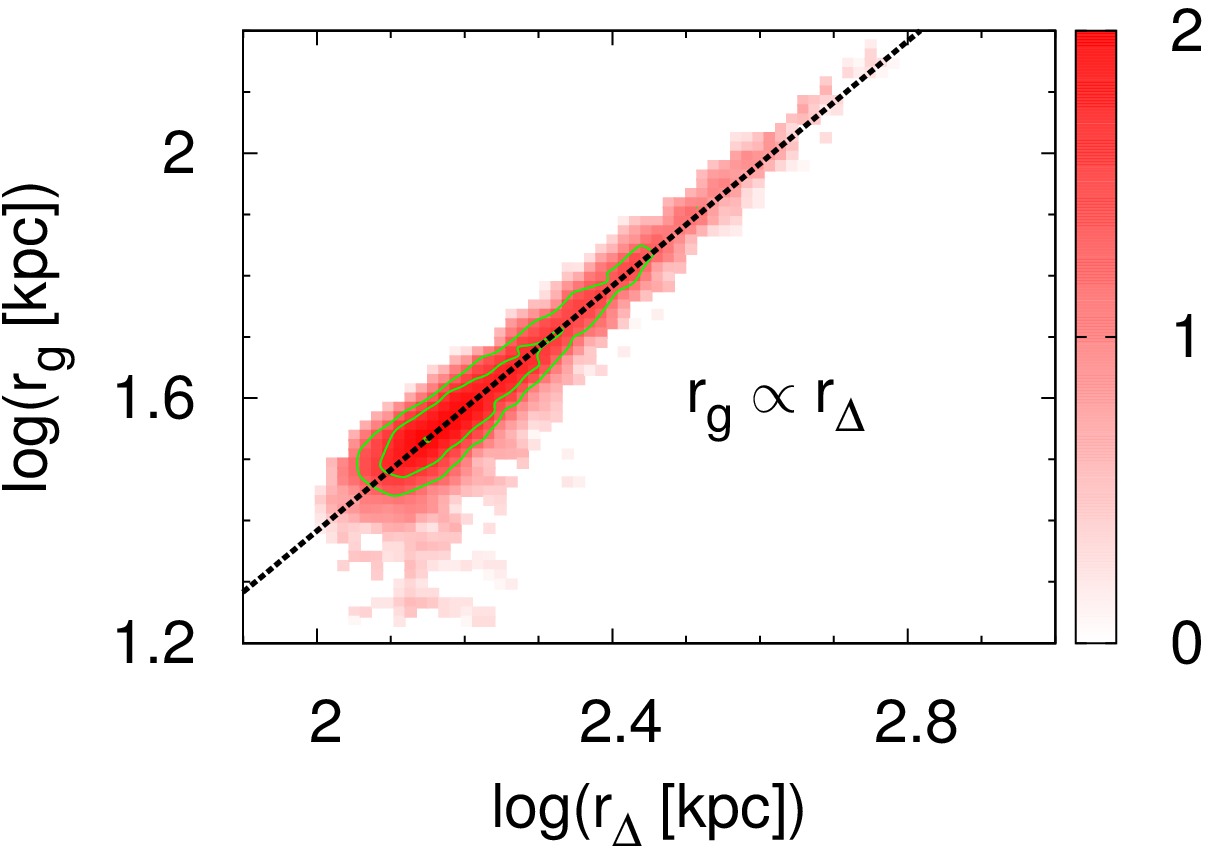}
\caption{Distributions of the simulated haloes at $z=0$ in the
  $\sigmazero-\sigmav$ (left-hand panel), $\rh-\rdelta$
  (central panel) and $\rg-\rdelta$ (right-hand panel) planes.  In each
  panel, the black short-dashed line is the \emph{linear} best-fit to
  the distribution.}
\label{fig:rh-rg-rv}
\end{figure*}

\subsection{Definitions of the structural and kinematical properties of dark matter haloes}
\label{sec:definitions}
In this Section we define the fundamental structural and kinematical properties of the
simulated DM haloes, such as the mass $M$, the size $r$ and the velocity dispersion $\sigma$.
There is not a unique method to identify a DM halo in a cosmological $N$-body simulation,
because, as for every non-truncated and non-isolated particle system, it is not trivial to
define the set of particles that belong to the object: various techniques have been
proposed in the astrophysical literature \citep[e.g., friends-of-friends or spherical overdensity
algorithms\footnote{
For a brief review of different methods see \cite{HgMAD}.
};
see][]{Efstathiou+1985,Davis+1985,LaceyCole1993,Evrard+2008}.
The problem is relevant, because different sets of particles can lead to different estimates
of the properties of the object.

In this work, we adopted the conventions of \cite{AHF} and we have used their \AHF finding code.
The haloes are identified from the peaks of a three dimensional density field calculated in a grid
with Adaptive Mesh Refinement (AMR).
The spherical region of radius $\rdelta$, which is centred at the centre of mass of the particles
in the highest refinement level of the AMR grid, defines the set of particles that belong to the
DM halo\footnote{
An \emph{unbinding procedure} is also run on the haloes in order to remove the gravitationally
unbound particles inside the spherical region \citep[see][]{AHF}.
} \citep[see][]{AHF}.
We use a standard definition of a halo as a certain spherical top-hat overdensity, via the formula
\begin{equation}
\label{def:halo}
\frac{3M}{4\pi \rdelta^3}=\Deltac(z)\rhoc(z),
\end{equation}
where $\rhoc (z) = 3 H^2(z)/8\pi G$ is the critical density of the Universe at redshift $z$, $\Deltac (z)$ is the
overdensity value at the same time, $H(z)$ is the Hubble parameter and $G$ is the gravitational constant. We adopt
the following definition of the critical overdensity in a flat Universe with negligible radiation energy density:
\begin{equation}
\label{def:Deltac}
\Deltac (z) = 18\pi^2 + 82\left[\Omega (z) -1\right] - 39\left[\Omega (z) -1\right]^2,
\end{equation}
where $\Omega(z)=\Omegazerom (1+z)^3/E(z)^2$ and
$E(z)^2=\Omegazerom(1+z)^3 + \Omegazerolambda$, such that the haloes
identified are expected to be close to equilibrium
\citep[see][]{LaceyCole1993, BryanNorman1998}.

Given the velocities of all particles belonging to the halo, we compute
the \emph{virial} velocity dispersion of the system
\begin{equation}
\label{def:sigmav}
\sigmav = \left[ \sum_{i=1}^N \sum_{j=1}^3 \frac{(v_{i,j} - \barv_{j})^2}{N} \right]^{1/2},
\end{equation}
where $N$ is the number of particles in the halo, $v_{i,j}$ is the $j$-th component of the $i$-th particle's velocity
and $\barv_{j}$ is the $j$-th component of the average velocity. 

The definition of the halo size is also non-trivial, since it is intimately affected by the choice
of the spherical overdensity. As a matter of fact the definition of $\rdelta$ via equation \eqref{def:halo}
is based on the idea that the material surrounding the overdensity is bound to the halo if the dynamical time
of the particle is less than the Hubble time at that redshift, so one can reasonably expect the halo of size
$\rdelta$ to be in equilibrium.
For this reason, $\rdelta$ is often, but improperly, called \emph{virial radius} and, in any case, in
the following we will adopt such convention. However, other definitions of the halo size are possible:
for example, the gravitational radius (i.e., the true virial radius)
\begin{equation}
\label{def:rg}
\rg \equiv \frac{GM}{\sigmav^2}
\end{equation}
and the half-mass radius $\rh$, i.e., the radius of the sphere enclosing half the total mass of the halo,
$M(<\rh)=M/2$.

We also employ a definition of the halo velocity dispersion alternative to $\sigmav$: we compute
the velocity dispersion profile $\sigma(<r)$ in the same fashion as the mass profile,
where the velocity dispersion is calculated as in equation \eqref{def:sigmav} and the sum is only on the particles within the
spherical region of radius $r$. From the $\sigma(<r)$ profile,
we then estimate the \emph{central velocity dispersion}
\begin{equation}
\label{def:sigma0}
\sigmazero \equiv \sigma\left(<\rh\right).
\end{equation}

We define our sample of DM haloes
by selecting those for which $\rh > 2\rconv$, where $\rconv$ is
the convergence radius in the sense of \cite{Power+2003}, i.e., the
radius within which the two-body collisions dominate the orbital
motions of the particles integrated by the code and the density
estimates are therefore unreliable. We verified that, for our
  sample, the mass contained within the convergence radius is
  typically a small fraction of that contained within the half-mass
  radius, so our measurement of $\sigma_0$ should be robust.  The
  sample selected with the above criterion is made of $\approx 11000$ DM
  haloes, with a lower limit in mass $\Mlower \simeq 1.3 \times
  10^{11}\Msol$.

\begin{figure*}
\includegraphics[width=0.49\hsize]{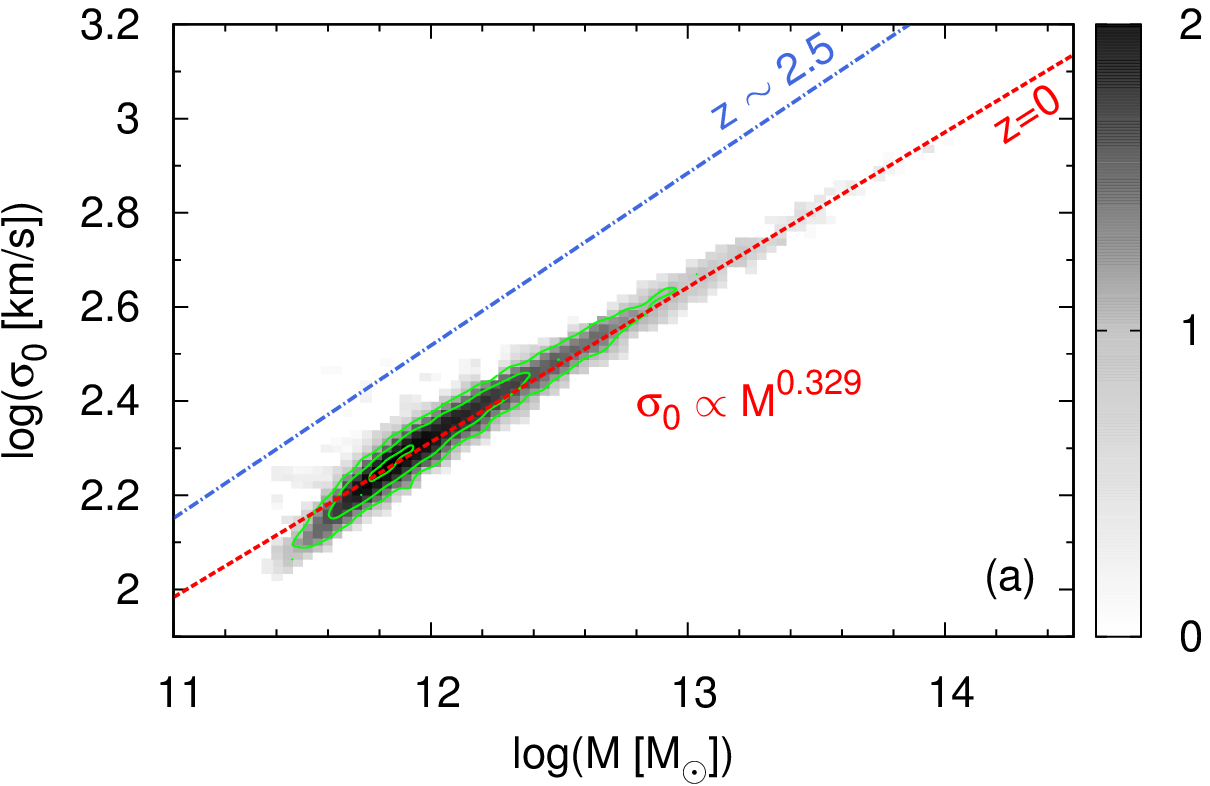}
\includegraphics[width=0.49\hsize]{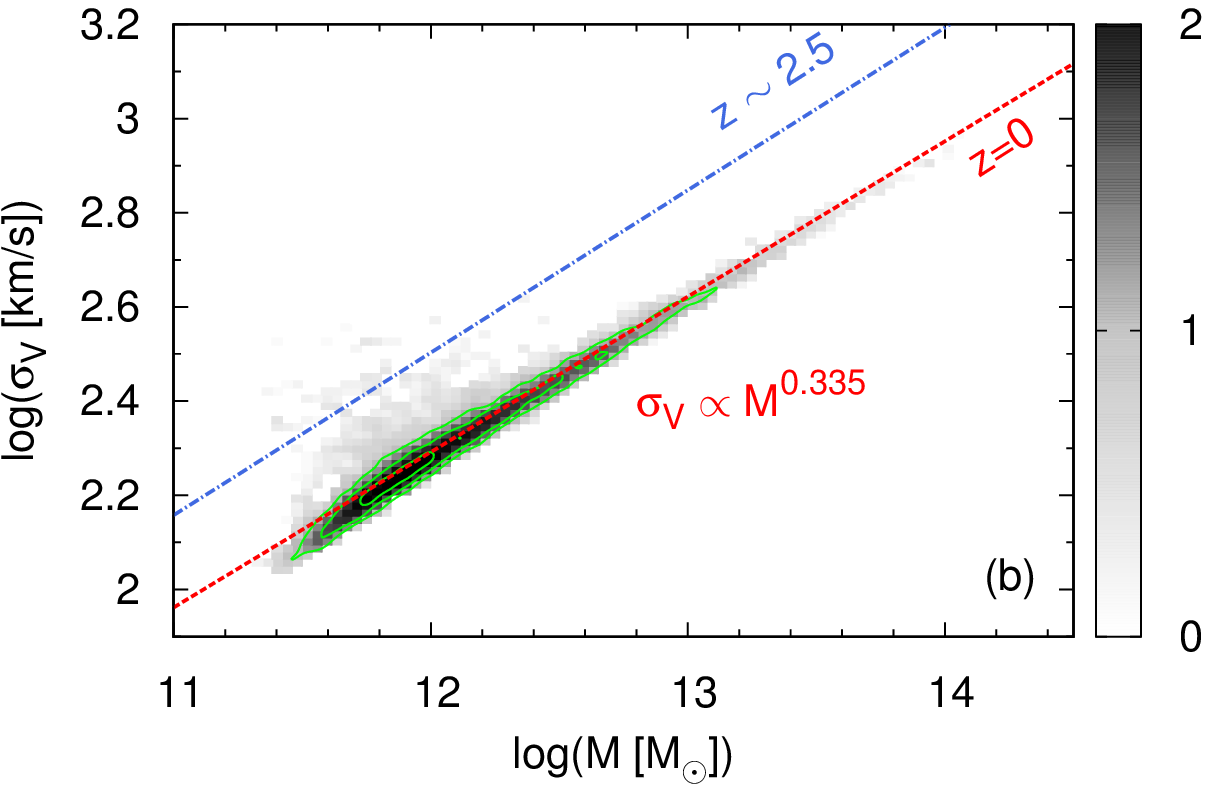}
\includegraphics[width=0.49\hsize]{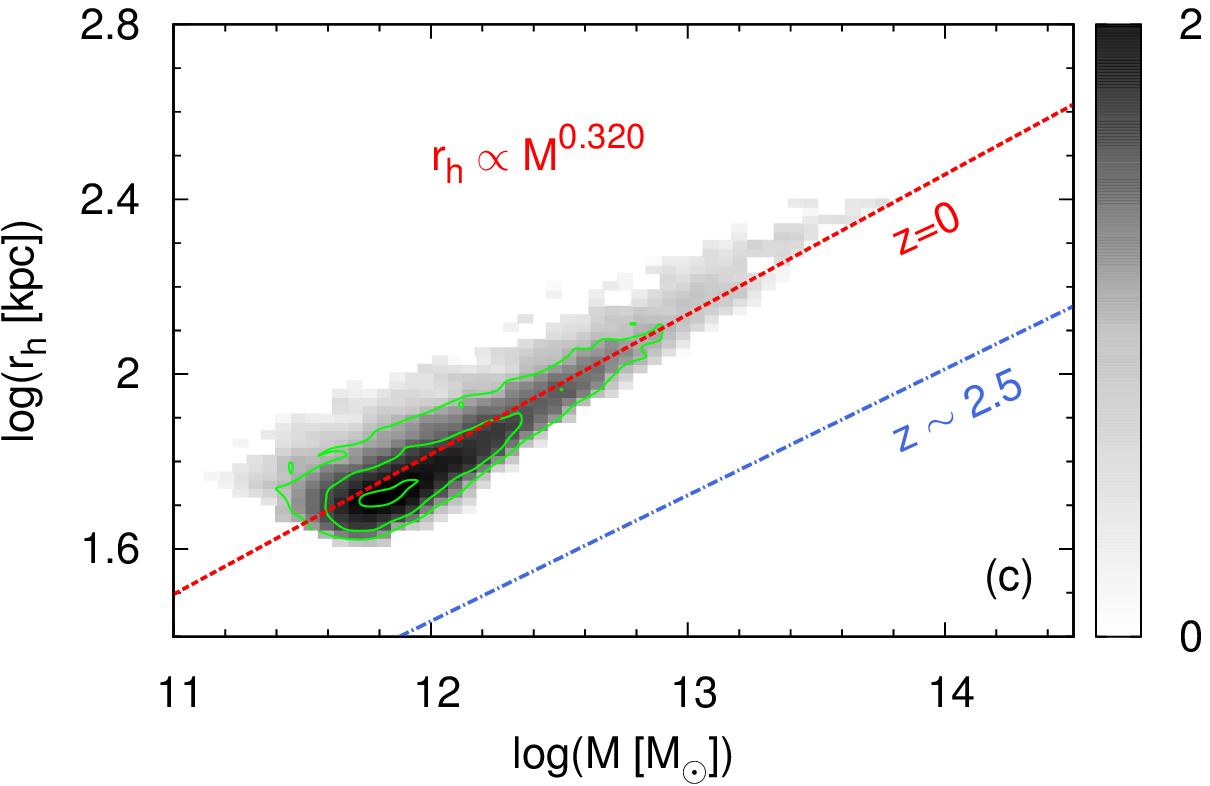}
\includegraphics[width=0.49\hsize]{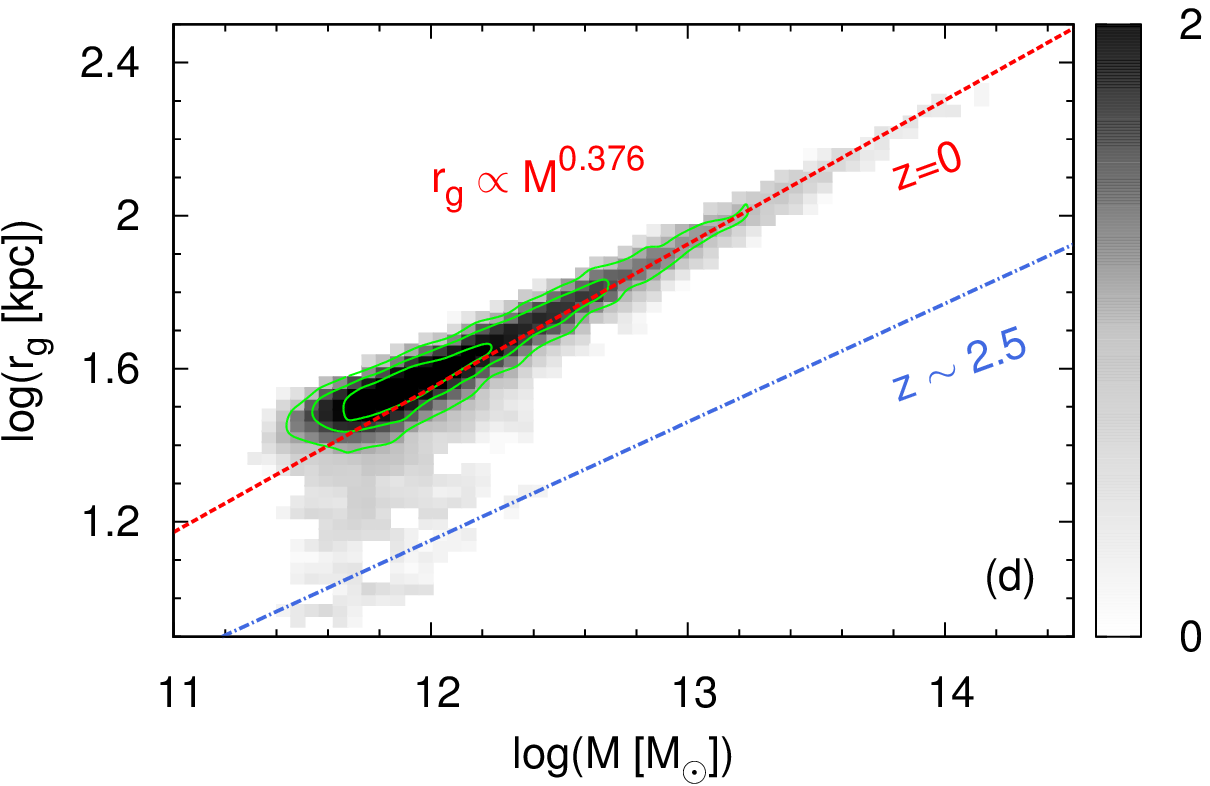}
\caption{Distributions at $z=0$ of the simulated haloes in the planes
  $M-\sigmazero$ (panel a), $M-\sigmav$ (panel b), $M-\rh$ (panel c)
  and $M-\rg$ (panel d). In each panel, the plane has been binned
    in cells and the gray scale represents the logarithm of the counts
    of haloes in each cell. The red dotted lines are the best-fits at
    $z=0$ ( $\sigmazero \propto M^{0.329}$, $\sigmav \propto
    M^{0.335}$, $\rh \propto M^{0.320}$ and $\rg \propto M^{0.376}$);
    for comparison, the best-fits at $z\simeq 2.5$ are plotted as blue
    dot-dashed lines.}
\label{fig:map_Msig0_Msigv_Mrh_Mrg}
\end{figure*}

\subsection{Behaviour of the different size and velocity dispersion proxies}
\label{sec:behaviour_diffdef}

In this section we discuss the relations between different size and
velocity dispersion estimators, in particular how the \emph{virial}
proxies ($\rdelta$, $\sigmav$) compare to the \emph{central} ones
($\rh$, $\sigmazero$). 

  In this work we use $\sigmazero$ and $\rh$ (see Section
  \ref{sec:definitions}) to characterize the haloes because it is
  reasonable to expect that the stellar central velocity dispersion
  $\sigmastar$ and the effective radius $\re$ of the galaxies
  inside such haloes are more related to the halo central quantities
  than to the halo virial quantities. However, different choices would
  be possible.
  For instance, a tight correlation between $\rdelta$ and $\re$ has been
  suggested, both theoretically \citep[see][]{Mo+1998} and using
  abundance matching techniques \citep[see][]{Kravtsov2013}. In
  addition, it is well known that the ratio $\rg/\rh$ depends
  only weakly on the density profile
  \citep[see][]{BT2008,Ciotti1991,Nipoti+2003}. It is therefore
  interesting to verify how the different proxies correlate in our
  sample of haloes and how their values depend on the choice of the
  critical overdensity $\Deltac$.

In Fig. \ref{fig:size-delta} we show the dependence of $\rdelta$,
$\rg$ and $\rh$ on $\Deltac$ at $z=0$ for a representative halo with
$M\simeq10^{13}\Msol$.
As expected from equation \eqref{def:halo}, the virial radius has a
perfect $\rdelta\propto\Deltac^{-1/3}$ scaling.  We note that also the
gravitational radius scales roughly as $\rg\propto\Deltac^{-1/3}$ and
this is because it depends only on virial quantities (see
equation \ref{def:rg}), while the half-mass radius has slightly less
steep dependence on $\Deltac$, namely $\rh\propto\Deltac^{-0.22}$.
This is an additional reason to prefer $\rh$ to $\rdelta$ in the
present context.  Formally, also the quantity $\sigmav$, and so
$\sigmazero$, depends on $\Deltac$, but the dependence is weak: the
variation is no more than $10\%$ in $\sigmav$, when varying $\Deltac$
by an order of magnitude.

In Fig. \ref{fig:rh-rg-rv} we show the distributions of the halo
population at $z=0$ in the $\sigmazero-\sigmav$, $\rh-\rdelta$ and
$\rg-\rdelta$ planes.  We find that in all cases a linear correlation
is in good agreement with the
distribution of the haloes in the three planes; fitting with
power laws we get $\sigmazero \propto \sigmav^{0.97 \pm 0.01}$,
$\rh\propto\rdelta^{0.96 \pm 0.01}$ and $\rg\propto\rdelta^{1.1 \pm
  0.01}$. Interestingly, for a given halo, the ratio
$\sigmazero/\sigmav$ is very close to unity: the average ratio in our
sample is $\langle\sigmazero/\sigmav\rangle = 1.01$. We recall
  that it is not a priori expected that the central quantities
  correlate linearly with the virial ones.  This finding indicates
  that the DM haloes are not systematically
  non-homologous: in other words, more massive haloes are, on
  average, just rescaled versions of less massive haloes (at least, as
  far as the relation between virial and central quantities is
  concerned).  The scatter around the linear relations in
  Fig. \ref{fig:rh-rg-rv} can be ascribed to some degree of
  non-homology at a given halo mass or to the fact that some haloes
  are not completely virialized.
  We note also that \cite{Diemer+2013b} recently found that there
  is a remarkable homology in their cluster-sized haloes sample:
  they argued that a tight relation exists between the mass and
  velocity dispersion profiles of DM haloes. Moreover, they claimed
  that the mass-velocity dispersion relation of the halo sample
  is almost insensitive to the size definition (in the range
  $100 < \Deltac < 2500$ in equation \ref{def:halo}) because of
  such homology in the radial profiles.
  
For the purposes of this work, given that, on average, the central
quantities ($\sigmazero$,$\rh$) scale linearly with the virial
quantities ($\sigmav$,$\rdelta$), our results would be virtually
unchanged if we adopted $\sigmav$, instead of $\sigmazero$, and
$\rdelta$ or $\rg$, instead of $\rh$, to characterize our haloes.

\subsection{Virial expectations for the halo mass-velocity dispersion and mass-size relations}
\label{sec:exp_Msig-Mr}

With the adopted halo definition (equation \ref{def:Deltac}) we
expect the DM haloes to be virialized in every
snapshot of the simulation from $z\simeq 2.5$ to $z \simeq 0$.
It follows that, some well known scaling laws are expected for the
dark-halo population: from equation (\ref{def:rg}), under the
assumption of a linear proportionality between the virial
radius (equation \ref{def:halo}) and the gravitational radius
(equation \ref{def:rg}) it follows \citep[e.g.,][]
{Lanzoni+2004} that for haloes in equilibrium
\begin{equation}
\label{eq:Msig3}
M\propto \sigmav^3.
\end{equation}
The correlation is expected to depend on redshift as
\begin{equation}
\label{eq:exp_z_Msig}
\sigmav \propto \left[ E(z) \,M\right]^{1/3}
\end{equation}
\citep[see e.g.,][]{Evrard+2008}. Using equation \eqref{def:halo}
to define the haloes and the definition of the critical density
$\rhoc=3 H^2/8\pi G$, for a flat Universe we get
\begin{equation}
\label{eq:expevo_rdelta1}
\rdelta = \left[ 2GM \, \Deltac(z)^{-1}\, H(z)^{-2}\right]^{1/3},
\end{equation}
where $H(z)=\Hubble E(z)$.
Assuming a linear dependence of the form $\rdelta=\xi\rg$, where $\xi$ is a dimensionless constant, then the virial
velocity dispersion can be written as $\sigmav^2=GM\xi/\rdelta$, implying
\begin{equation}
\label{eq:sigmav(M,z)}
\sigmav = \xi^{1/2} \left( \frac{GM}{\sqrt{2}}\right)^{1/3}\, \Deltac(z)^{1/6} H(z)^{1/3}.
\end{equation}
We have seen in Section \ref{sec:behaviour_diffdef} that
$\sigmazero\propto\sigmav$, so also the mass and redshift dependence of
$\sigmazero$ is expected to be given by equation
\eqref{eq:sigmav(M,z)}.
Under the assumption that $\rh$ and $\rg$ scale linearly with
$\rdelta$ (see Section~\ref{sec:behaviour_diffdef}), from
equation~(\ref{eq:expevo_rdelta1}) we have that a fixed $z$
\begin{equation}
\label{eq:Mrg3}
\rh\propto\rg \propto M^{1/3}.
\end{equation}
The $z$-dependence of the other size proxies are also given by equation \eqref{eq:expevo_rdelta1}, as $\rg\propto\rdelta$ and
$\rh\propto\rdelta$ (see Section \ref{sec:behaviour_diffdef}).

\section{Scaling relations of dark matter haloes as functions of redshift}
\label{sec:tot_scaling}

Here we present the distributions of the size and the velocity
dispersion as functions of mass and redshift for our sample of $\sim
11000$ DM haloes (see Section \ref{sec:definitions}) in the mass range
$10^{11}\Msol \leq M \leq 5.5 \times 10^{14}\Msol$.  We adopt the same
lower mass limit in all the snapshots of our simulation, while we do
not restrict the upper mass limit, which varies from $M \simeq 2.67
\times 10^{13}\Msol$ at $z \simeq 2.5$ to $M \simeq 5.5 \times
10^{14}\Msol$ at $z=0$.

\begin{figure}
\includegraphics[width=\hsize]{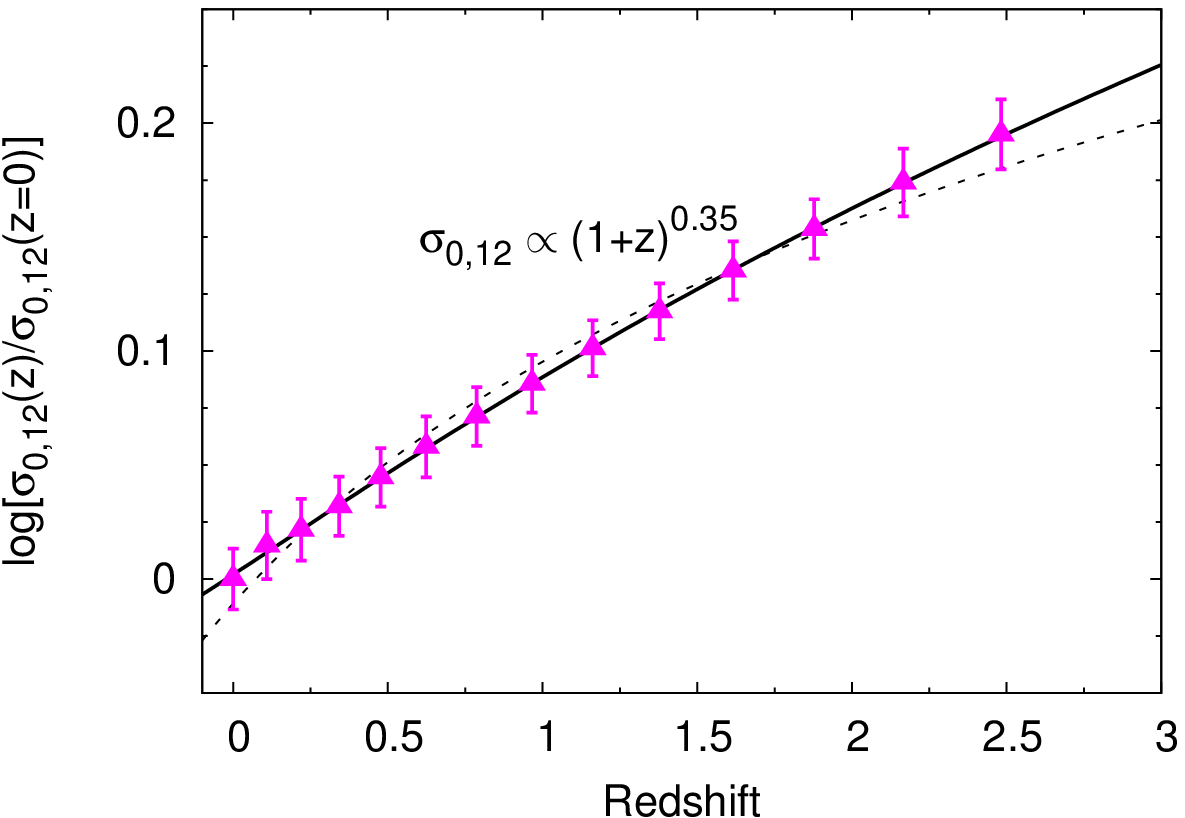}
\includegraphics[width=\hsize]{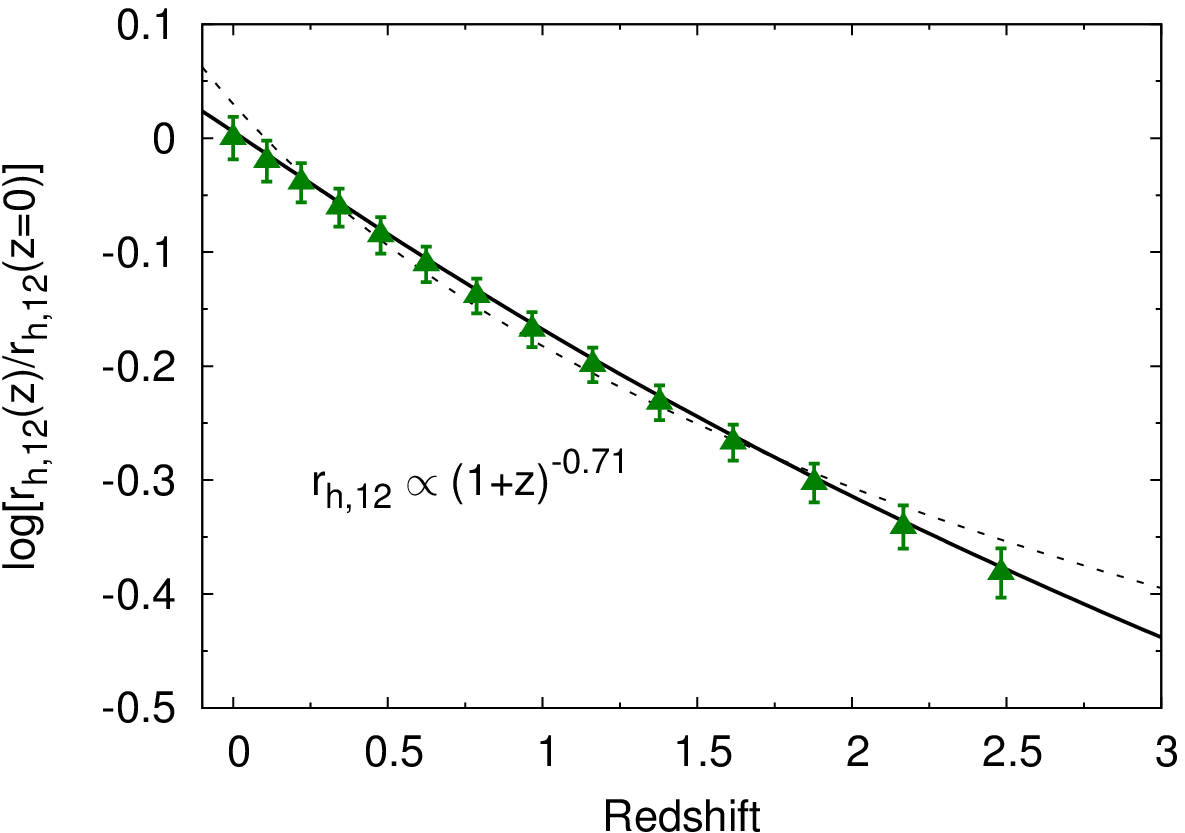}
\caption{Top panel: halo central velocity dispersion at fixed $M=10^{12}\Msol$ as a function of
	  redshift. The black dashed line is the power-law best-fit to the points,
	  while the black solid line is a fit quadratic in logarithm (see text).
	  Bottom panel: same as top panel but for the half-mass radius at fixed $M=10^{12}\Msol$.
}
\label{fig:sig0_12-rh_12}
\end{figure}

\subsection{Mass-velocity dispersion: the measured correlation and evolution}
\label{sec:Msig}
The simulated DM halo population is well represented by the following best-fit relation:
\begin{equation}
\label{eq:bestfit_Msigzero}
\log\left( \frac{\sigmazero}{\rm km/s} \right) = \gamma \log\left( \frac{M}{\Msol}\right) + b,
\end{equation}
where
\begin{equation}
\begin{split}
 \log\gamma& = (0.103 \pm 0.008) \log(1+z) - (0.49 \pm 0.002), \\
 \log b& = (0.163 \pm 0.02) \log(1+z) + (0.179 \pm 0.006).
\end{split}
\end{equation}
In Fig. \ref{fig:map_Msig0_Msigv_Mrh_Mrg}(a) we show the distribution
of the sample in the $M-\sigmazero$ plane at $z=0$ and we find a
best-fit correlation of the type $M \propto \sigmazero^{3.04 \pm 0.01}$
(i.e. $\gamma\simeq 0.32$).  For comparison we also show the best-fit
correlation at $z \simeq 2.5$: the slope is slightly larger than that
at $z=0$, i.e., $\gamma \simeq 0.37$, whereas the normalization is
significantly higher: as we expected, at fixed mass higher $z$
haloes have higher velocity dispersion.

Assuming $\sigmazero\propto\sigmav$ (see Section
\ref{sec:behaviour_diffdef}), we can compare our results with both
theoretical expectations, e.g., equation \eqref{eq:Msig3}, and
previous findings.  At fixed mass $M=10^{12}\Msol$ we find a good
agreement with equation \eqref{eq:exp_z_Msig}: i.e., our sample
follows\footnote{ We define $A_{\rm X,Y} \equiv A_{\rm
    X}(M=10^{\rm Y}\Msol)$.}  $\sigma_{0,12} \propto E(z)^{0.36}$.
In the top panel of Fig. \ref{fig:sig0_12-rh_12} we show the
evolution of the central velocity dispersion at $M=10^{12}\Msol$ in
the redshift range $0\lesssim z \lesssim 2.5$. We have chosen $M=10^{12}
\Msol$ as a reference mass, since it is
roughly the mean mass in our sample at $z=0$ and it is still
well resolved (about $8\times 10^2$ particles). As expected from
cosmological predictions (equation \ref{eq:exp_z_Msig}), 
$\sigma_{0,12}$ decreases with time.  We find a power-law best-fit
evolution of the type $\sigma_{0,12} \propto (1+z)^{0.35}$
and also a better representation (about two orders of magnitude in the
reduced $\chi^2$) of the results via the fitting formula
$\log\sigmazero = 0.29\, x^2+0.2\, x$, where $x\equiv\log(1+z)$.
We comment here that fixing a typical mass, say $M=10^{12}\Msol$ as
in Fig. \ref{fig:sig0_12-rh_12}, means that we are analysing different
haloes at different $z$, unlike fixing a halo and focusing on its
evolution.

To further compare the results of our simulation with theoretical
predictions and previous works, we have analysed the correlation
between the mass and the virial velocity dispersion $\sigmav$. In
Fig. \ref{fig:map_Msig0_Msigv_Mrh_Mrg}(b) we plot the distribution of
the DM haloes in the $M-\sigmav$ plane.  In general, there is a very
good agreement with the theoretical expectation \eqref{eq:Msig3}
derived from the equilibrium assumption: the best-fit relation
corresponds to $M\propto\sigmav^{2.97 \pm 0.01}$.  Other authors
found similar results from independent simulations \citep[see
  e.g.,][]{Evrard+2008, Stanek+2010,Munari+2013,Diemer+2013b}.
We fitted the evolution of the normalization, at $M=10^{12} \Msol$,
of the $M-\sigmav$ correlation for our simulated haloes as a function
of $E(z)$.  We find that from
$z\simeq 2.5$ to $z \simeq 0$ the normalization at $M=10^{12} \Msol$ follows
$\sigmatwelve \propto E(z)^{0.33}$, which is in remarkably
good agreement with theoretical expectations (equation
\ref{eq:exp_z_Msig}).
  We compare also our results with previous findings: for $M=10^{14}
  \Msol$ \cite{Stanek+2010} estimate an evolution $\sigma_{\rm V,14}
  \propto E(z)^{0.34}$ and we find $\sigma_{\rm V,14} \propto
  E(z)^{0.35}$; for $M=10^{14.3} \Msol$ \cite{Lau+2010} find
  $\sigma_{\rm V,14.3} \propto (1+z)^{0.49}$ in their non radiative
  case (with fixed $\Deltac=500$) and we find $\sigma_{\rm V,14.3}
  \propto (1+z)^{0.36}$.

\subsection{Mass-size: the measured correlation and evolution}
\label{sec:Mr}
The simulated DM halo population is well represented by the following best-fit relation:
\begin{equation}
\label{eq:bestfit_Mrh}
\log\left(\frac{\rh}{\rm kpc}\right) = \gamma \log\left(\frac{M}{\Msol}\right) + b,
\end{equation}
where
\begin{equation}
\begin{split}
 \log\gamma& = (-0.069 \pm 0.01) \log(1+z) + (-0.489 \pm 0.003), \\
 \log b& = (0.02 \pm 0.02) \log(1+z) + (0.312 \pm 0.004).
\end{split}
\end{equation}
The measured correlation for the simulated haloes is shown in Fig. \ref{fig:map_Msig0_Msigv_Mrh_Mrg}(c).
The best-fit relation computed for this sample at $z=0$ is $M \propto \rh^{3.12 \pm 0.02}$ (i.e. $\gamma\simeq 0.32$).
For comparison, we also plot here the $z \simeq 2.5$ best-fit
correlation: the slope is slightly decreasing with redshift, down to
$\gamma \simeq 0.28$ and the normalization in this
mass range gets lower at later times. As we expected, at fixed mass
higher $z$ haloes have smaller size, i.e., they have higher density.

  We then fit the evolution of the normalization at
  $M=10^{12}\Msol$ of the $M-\rh$ correlation as a function of
  redshift in the range $0<z<2.5$: we find an evolution of the type
  $\rhtwelve \propto E(z)^{-0.65}$, which is in good
  agreement with the expectations given by equation
  \eqref{eq:expevo_rdelta1}. In the bottom panel of
  Fig. \ref{fig:sig0_12-rh_12} our findings on the evolution in time
  of $\rhtwelve$ are shown: as expected
  (equation \ref{eq:expevo_rdelta1}), we find that $\rhtwelve$
  increases with time.  We find a power-law best-fit
  $\rh\propto(1+z)^{-0.71}$ and also that a better representation of
  the results is given by $\log\rh=-0.53\, x^2-0.41\, x$, where
  $x\equiv\log(1+z)$.
  
\begin{figure}
\includegraphics[width=\hsize]{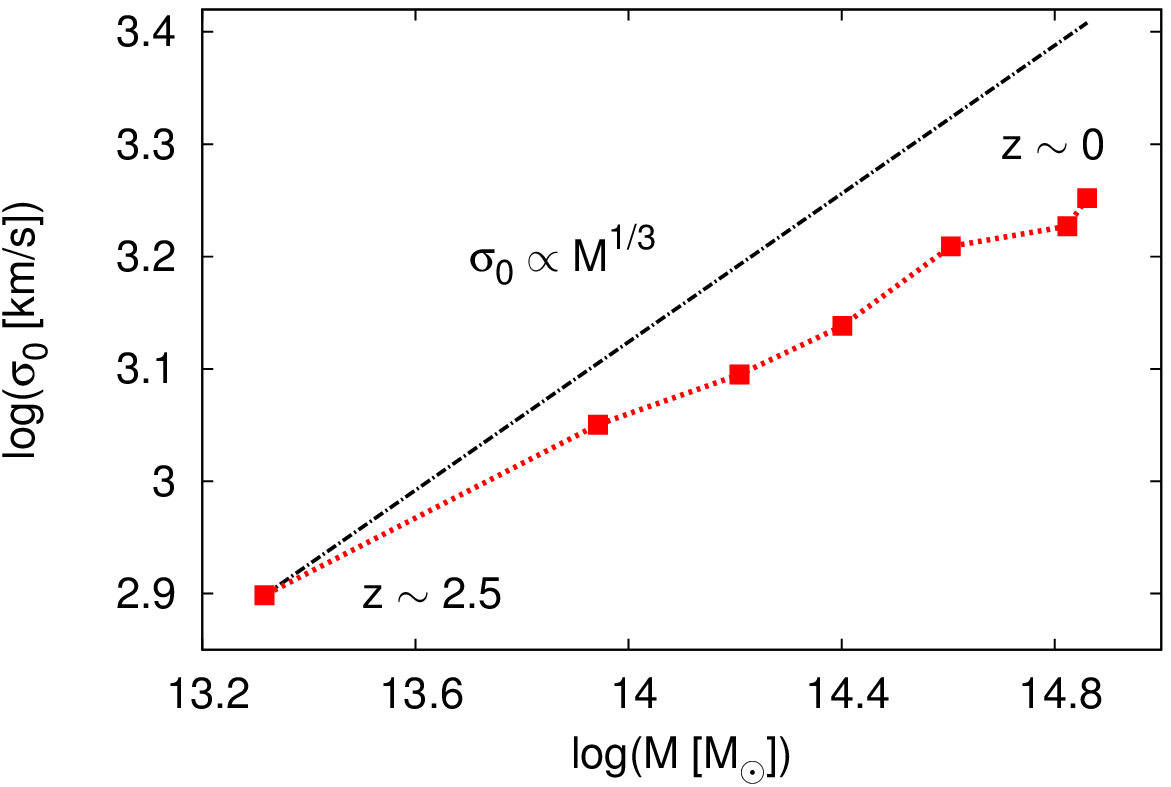}
\includegraphics[width=\hsize]{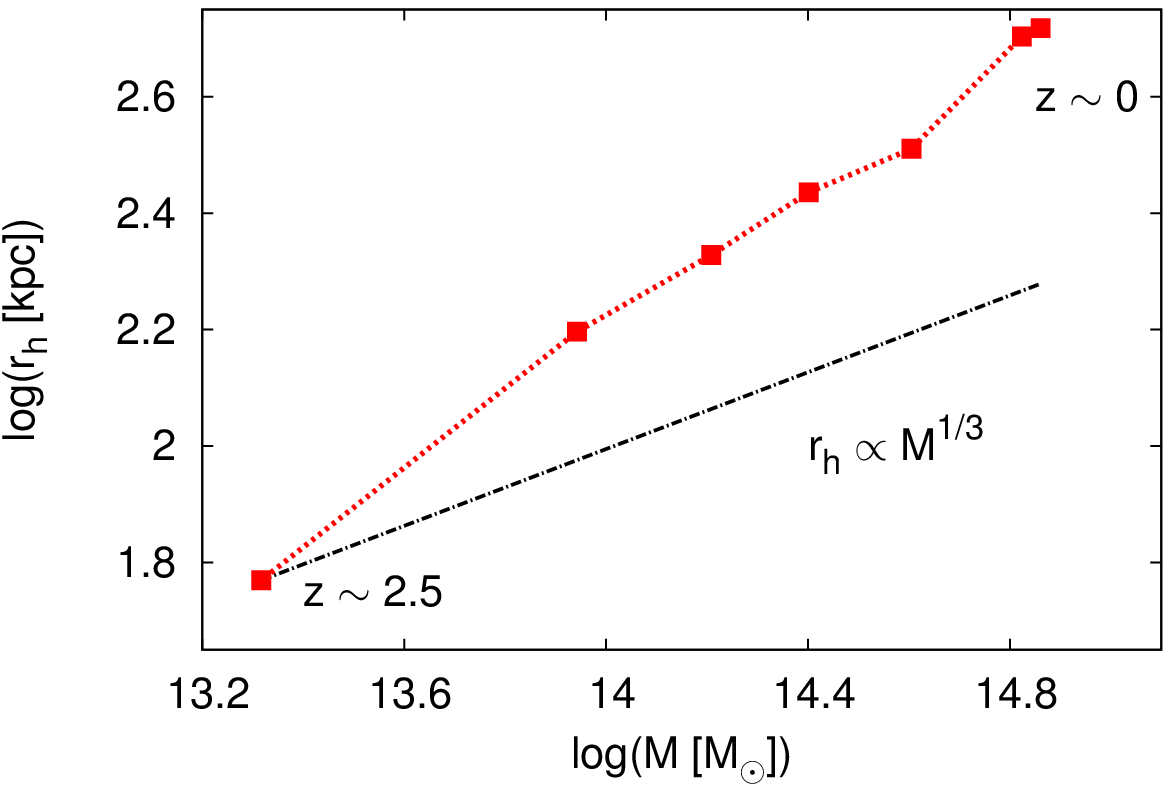}
\caption{ Top panel: evolution of a DM halo of mass $M(z=0) \simeq 5.5
  \times 10^{14} \Msol$ in the $M-\sigmazero$ plane from $z\simeq 2.5$
  to $z=0$ (red squares). Here, mass traces time: the least massive
  point is the one at highest $z$. The black dot-dashed line is a
  power law of index $1/3$, plotted for comparison. Bottom panel: same
  as top panel, but in the $M-\rh$ plane.}
\label{fig:evo_Msig_Mrh}
\end{figure}

\begin{figure}
\includegraphics[width=\hsize]{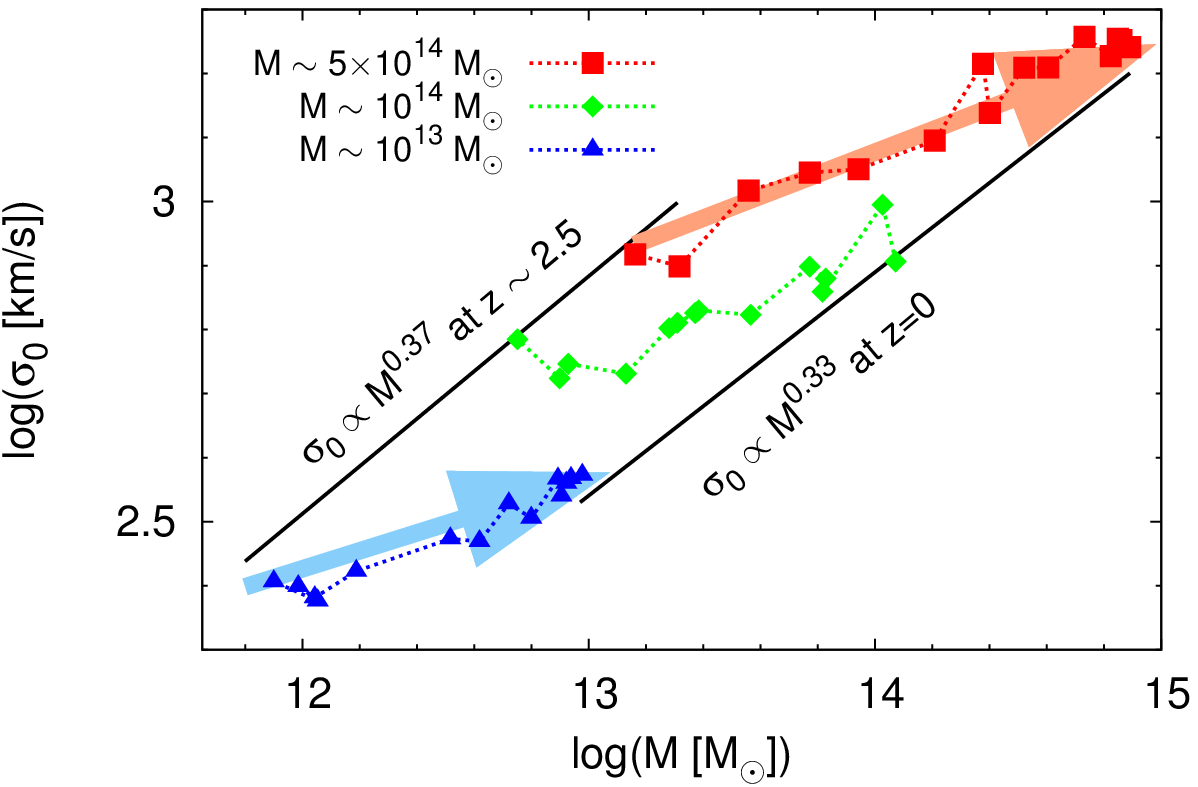}
\includegraphics[width=\hsize]{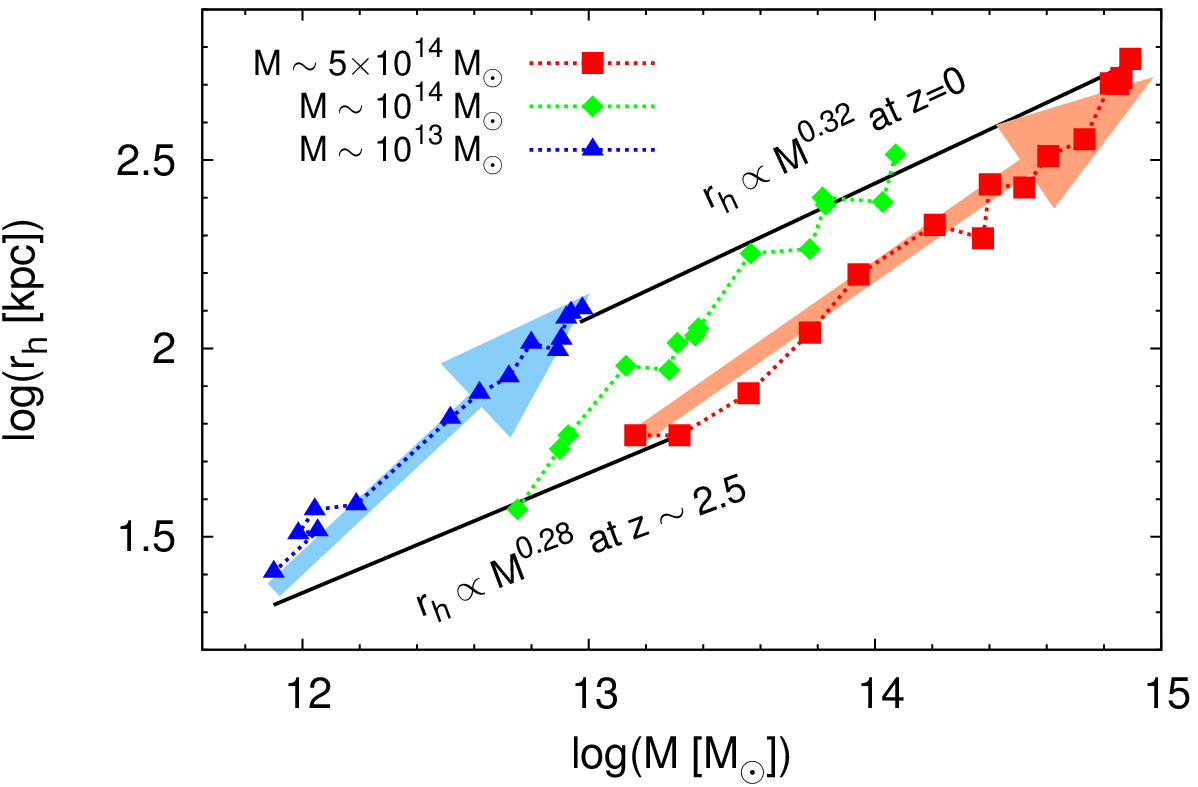}
\caption{Top panel: evolution in the $M-\sigmazero$ plane of three
  representative haloes in our simulation. The red filled squares
  refer to a cluster-sized halo of mass $M(z=0) \simeq 5.5 \times
  10^{14}\Msol$ (the same halo as in Fig. \ref{fig:evo_Msig_Mrh},
  but with a finer sampling), the green filled diamonds refer to a
  group-sized halo of mass $M(z=0) \simeq 10^{14}\Msol$, while the blue
  filled triangles refer to a early-type galaxy-sized halo of mass $M(z=0)
  \simeq 10^{13}\Msol$. The two black solid lines are the best-fit
  $M-\sigmazero$ correlation for the whole population at $z=0$ and at
  $z \simeq 2.5$.  The big blue and red arrows represent a schematic
  view of the evolutionary track in the $M-\sigmazero$ plane for,
  respectively, the low-mass and high-mass haloes.  Bottom panel: same
  as top panel, but in the $M-\rh$ plane.}
\label{fig:wholevo_Msig_Mrh}
\end{figure}

In Fig. \ref{fig:map_Msig0_Msigv_Mrh_Mrg}(d) we plot the distribution
of the DM haloes in the $M-\rg$ plane.  Also in this case we find a
reasonably good agreement with the virial expectation \eqref{eq:Mrg3},
with a best-fit correlation $M \propto \rg^{2.71 \pm 0.02}$.  We
notice here that the somehow steeper slope than expected in
Fig. \ref{fig:map_Msig0_Msigv_Mrh_Mrg}(d) is due to a
\emph{tail} of the distribution composed of low mass haloes
($M<10^{12}\Msol$) with small $\rg$:
this feature can be due to the fact that some DM haloes are not in equilibrium, for instance, because they could have
experienced a recent major merger, and so virialization is not a good assumption for such objects.

\section{Evolution of individual haloes}
\label{sec:evo_sing}
\subsection{Evolution of simulated dark haloes in the mass-velocity dispersion and mass-size planes}
\label{sec:singlevo_Msig-Mrh}

According to the halo definition here adopted (equation \ref{def:halo})
the halo mass increases monotonically with time,
so studying a property of a halo as a function of its mass is
equivalent to studying the time-evolution of the same property.
Here we present the time-evolution of $\sigmazero$ and $\rh$ for
some individual representative haloes in our simulation, by tracking
them in the planes $\sigmazero-M$ and $\rh-M$. In
Fig. \ref{fig:evo_Msig_Mrh} we plot the evolutionary tracks followed
by a representative halo with mass $M \simeq 5.5 \times 10^{14}\Msol$
at $z=0$: we reconstruct the growth in velocity dispersion (top panel)
and size (bottom panel) as the halo gets more massive.  It is apparent
that neither in the $M-\sigmazero$ nor in the $M-\rh$ plane the halo
evolves along the scaling law of slope $\approx 1/3$.
The actual evolution experienced by the halo in
Fig. \ref{fig:evo_Msig_Mrh} is significantly shallower in the
$M-\sigmazero$ plane, with best-fit $\sigmazero\propto M^{0.2}$,
and steeper in the $M-\rh$ plane, with best-fit $\rh\propto M^{0.6}$.,
consistent with the results of previous works on binary
dissipationless mergers \citep[see e.g., \citetalias{Nipoti+2003};][]{Boylan-Kolchin+2005,
Hopkins+2009a,Hilz+2012,Hilz+2013}.

Figure \ref{fig:wholevo_Msig_Mrh} gives an overall picture of the evolution
in the $M-\sigmazero$ (upper panel) and $M-\rh$ (lower panel) planes of the
whole halo population. In particular, we follow the evolution from $z \simeq 2.5$
of three objects, having $M \simeq 10^{13}\Msol$, $M \simeq 10^{14}\Msol$
and $M \simeq 5.5 \times 10^{14}\Msol$ at $z=0$, representative of an ETG sized
halo, a group sized halo and a cluster sized halo, respectively.
Figure \ref{fig:wholevo_Msig_Mrh} indicates that the evolutionary tracks
of the whole population of DM haloes in the simulation reflect that of the halo
shown in the of Fig. \ref{fig:evo_Msig_Mrh}, therefore we do not find
clear indications of mass-dependent evolution. In other words, the velocity
dispersion grows weakly and the half-mass radius grows strongly independently of
the halo mass.
The fact that the individual haloes experience an evolution in $\sigmazero$
with a shallower slope than the global $\sigmazero\propto M^{1/3}$ correlation
(and viceversa for the evolution in $\rh$) is responsible for the $z$-evolution
of the normalization of the $\sigmazero-M$ correlation, such that at fixed
mass $\sigmazero$ is smaller at lower $z$. Similarly, the $z$-evolution of
the normalization of the $\rh-M$ correlation is such that at fixed mass $\rh$
is larger at lower $z$.

\subsection{Comparison with simple dry merger models}
\label{sec:drymerg}

In a hierarchical context, the evolution and mass assembly of haloes
is often decomposed it two main processes: 
diffuse accretion and mergers \citep[e.g.][]{Fakhouri+2010}.
According to the halo definition here adopted (equation
\ref{def:halo}), the halo mass can grow in principle even in
an isolated and static configuration, just because the critical
density of the Universe decreases with time
\citep[see e.g.,][]{Diemer+2013a}. However, in a realistic
cosmological context a halo experiences several mergers in its
lifetime and in many cases they dominate its mass assembly.
In the following, we will consider the case in which merging is
the driving process for the structural evolution of individual
haloes and we will compare our results with predictions of simple
dry merging models.

\subsubsection{Analytic arguments}
\label{sec:merg_analytic}

Here we present some of the analytic arguments one can use to
describe the evolution of the velocity dispersion and size of a halo
which grows mainly via mergers with other haloes.  If both the mass
loss in the collision and the orbital energy of the encounter are
negligible, in an equal-mass merger scenario the halo is expected to
grow in size linearly with mass, while its velocity dispersion is
expected to remain constant \citepalias[see e.g.,][]{Nipoti+2003}. Under
the same hypothesis, if the evolution is dominated by accretion of
many satellites much less massive than the main halo, the velocity
dispersion is expected to \emph{decrease} linearly with the mass,
while the size would grow quadratically with mass \citep[see
  e.g.,][]{Naab+2009}.  In a realistic merging history consisting of
both major and minor mergers, we expect the behaviour of $\rh$ and
$\sigmazero$ to be in between those predicted by the two extreme cases
illustrated above.

However the assumptions of zero orbital energy and negligible mass
loss are not necessarily realistic.  Some authors have shown how the
effect of mass loss could play an important role in the evolution of
the velocity dispersion and of the size of an object \citep[see
  e.g., \citetalias{Nipoti+2003};][]{Hilz+2012}. Also the effect of the
  orbital energy on the evolution of $\sigmazero$ and $\rh$ can be
  non-negligible \citep[][hereafter \citetalias{Nipoti+2012}]
  {Nipoti+2009a,Nipoti+2012}, even if there
  are indications that most halo encounters are on orbits close to
  parabolic \citep[see e.g.,][]{KhochfarBurkert2006}. In this
section we attempt to study the effect of orbital energy using the
data of our simulation.

\begin{figure}
\includegraphics[width=\hsize]{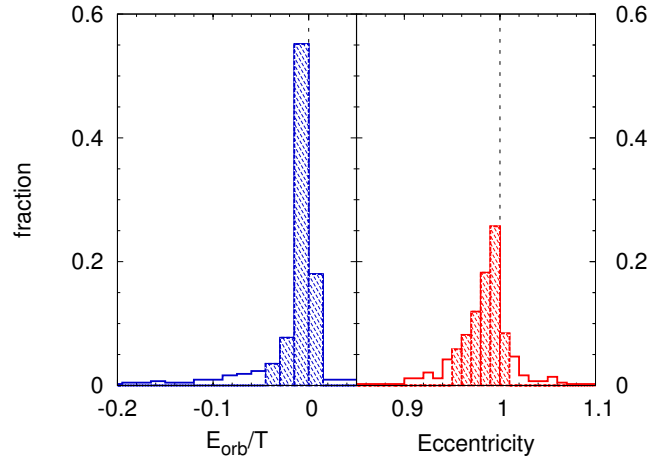}
\caption{Histograms of the orbital energies (left-hand panel) and of
  the eccentricities (right-hand panel), computed in the PMA, 
  of the mergers experienced by a $M(z=0)\simeq5.5\times 10^{14}\Msol$
  DM halo, from $z\simeq 2.5$ to $z=0$. The orbital energies are
  normalized to the internal energy $T=M\sigma_{V}^2/2$ of the
  reference halo. The dashed area, centred on the median value,
  contains $80\%$ of the counts, so that each tail of the distribution
  accounts only for $10\%$.}
\label{fig:eccentricity}
\end{figure}

\begin{figure*}
\includegraphics[width=\hsize]{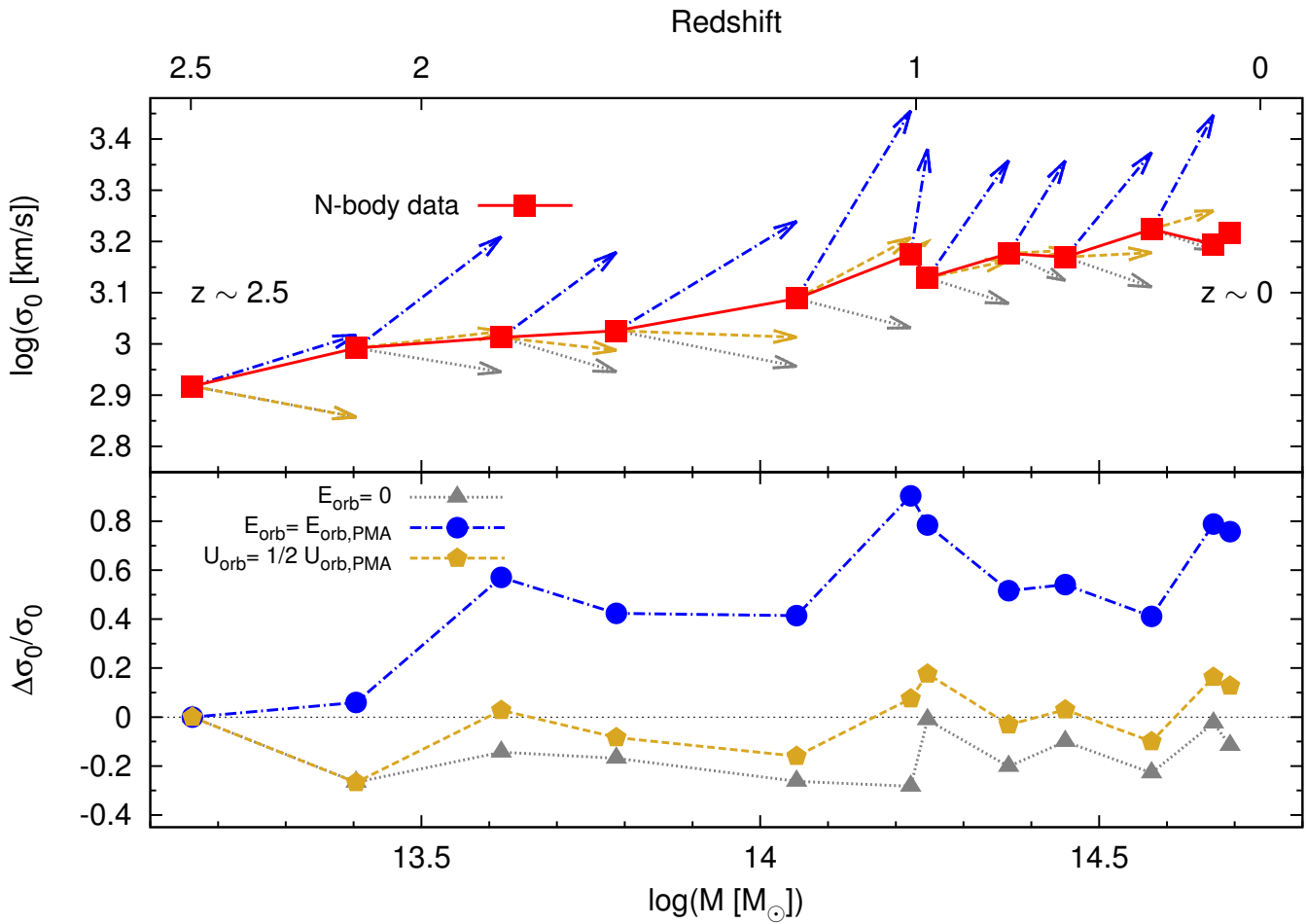}
\caption{Top panel: evolution of a $M(z=0)\simeq 5.5
  \times 10^{14}\Msol$ halo in the $M-\sigmazero$ plane from $z\simeq
  2.5$ to $z\simeq 0$ (red filled squares). The different arrows
  represent the predictions of the $\sigmazero$ in the different
  models, starting from the reference point which is
  $(M,\sigmazero)$ of the halo measured at the previous snapshot:
  gray dotted arrows are in the parabolic ($\Eorb=0$) model,
  the blue dot-dashed arrows in the $\Eorb = \EPMA$ model, in which
  the orbital energy is computed in the PMA,
  while the golden short-dashed arrows are in the $\Uorb = \UPMA/2$
  model, in which the orbital potential energy is half that in the PMA
  (see text). Bottom panel: relative deviations of the
  model predictions from the measured $N$-body data.}
\label{fig:evotracks_Msig}
\end{figure*}

  If we know the orbital energy, we can predict analytically the
  merger-driven evolution of the virial velocity dispersion $\sigmav$:
  the dissipationless merging of two virialized systems, which have
  kinetic energies respectively $T_1 = M_1 \sigma_{\rm V,1}^2/2$ and
  $T_2 = M_2 \sigma_{\rm V,2}^2/2$, on a barycentric orbit of energy
  $\Eorb$, results in a system that, when in equilibrium, has virial
  velocity dispersion \citepalias[see][]{Nipoti+2003}
\begin{equation}
\label{eq:sig_nonpar}
\sigmavf^2 = \frac{M_1\sigma_{\rm V,1}^2 + M_2\sigma_{\rm V,2}^2}{M_1+M_2} - 2\frac{\Eorb}{M_1+M_2},
\end{equation}
assuming no mass loss (the subscript f indicates the final value). As
long as $\sigmazero\propto\sigmav$ (see Section
\ref{sec:behaviour_diffdef}), equation \eqref{eq:sig_nonpar} can be
used to predict also the evolution of $\sigmazero$.

Similarly, one can predict how the size of the halo is evolving in the
merging process: assuming a linear proportionality $\rh\propto\rg$
(see Section \ref{sec:Mr} and Fig. \ref{fig:rh-rg-rv}), we have that
$\rh\propto M/\sigmav^{2}$. We can use
  equation~\eqref{eq:sig_nonpar} to make predictions in the context of
  a simple dry merging model: we calculate the half-mass radius of the
  halo at a given redshift, then we predict its evolution calculating
  the ratio
\begin{equation}
\label{eq:rh_nonpar}
\frac{\rhf}{\rhi}=\frac{M_1 + M_2}{M_1}\frac{\sigmavi^2}{\sigmavf^2},
\end{equation}
where $\sigmavf^2$ was calculated through equation \eqref{eq:sig_nonpar}
and we have taken halo $1$ as reference progenitor.

\subsubsection{Orbital parameters}
\label{sec:orbitalpar}

  To account for the contribution of $\Eorb$ in the evolution of
  $\sigmazero$ and $\rh$ for our haloes (i.e to apply
  equation~\ref{eq:sig_nonpar}), we need to extract the merger orbital
  parameters from our simulation. We have reconstructed the merger
histories of the haloes and then calculated the orbital parameters
of the encounters in the point-mass approximation (hereafter PMA) of
the progenitors \citep[i.e., approximating every merging halo as a
  point located in its centre of mass and having the same mass as the
  object; see e.g.,][]{KhochfarBurkert2006, Wetzel2011}.

As well known, the orbit of a collision is completely characterized by two
parameters, for instance the orbital energy and the orbital angular
momentum or the eccentricity and the pericentric radius. Here we
  find convenient to characterize the orbits of our mergers with the
  orbital energy $\Eorb$ and the eccentricity
\begin{equation}
\label{def:eccentricity}
e=\sqrt{1+\frac{2\Eorb \Lorb^2}{\mu(GM_1 M_2)^2}},
\end{equation}
where $M_1$ and $M_2$ are the masses of the two colliding systems,
$\Lorb$ is the norm of the barycentric orbital angular momentum
and $\mu \equiv M_1M_2/(M_1+M_2)$ is the reduced mass. In
Fig. \ref{fig:eccentricity} we plot the histograms of $\Eorb$ and $e$
(computed in the PMA) of the encounters
experienced by a $M(z=0)\simeq5.5\times 10^{14}\Msol$ DM halo in our
simulation from $z\simeq 2.5$ to $z=0$.  We find that the distribution
of the orbital energies has a clear peak at $\Eorb \simeq 0$ (left-hand panel
of Fig. \ref{fig:eccentricity}) and that of the eccentricities has a
clear peak at $e \simeq 1$ (right-hand panel of Fig. \ref{fig:eccentricity}),
with both distributions having non-negligible tails both at bound
orbits ($e<1$ and $\Eorb<0$) and unbound orbits ($e>1$ and $\Eorb>0$).
Overall, our findings are in agreement with those of
\cite{KhochfarBurkert2006}: the large majority of the mergers happen
on orbits close to parabolic.

\begin{figure*}
\includegraphics[width=\hsize]{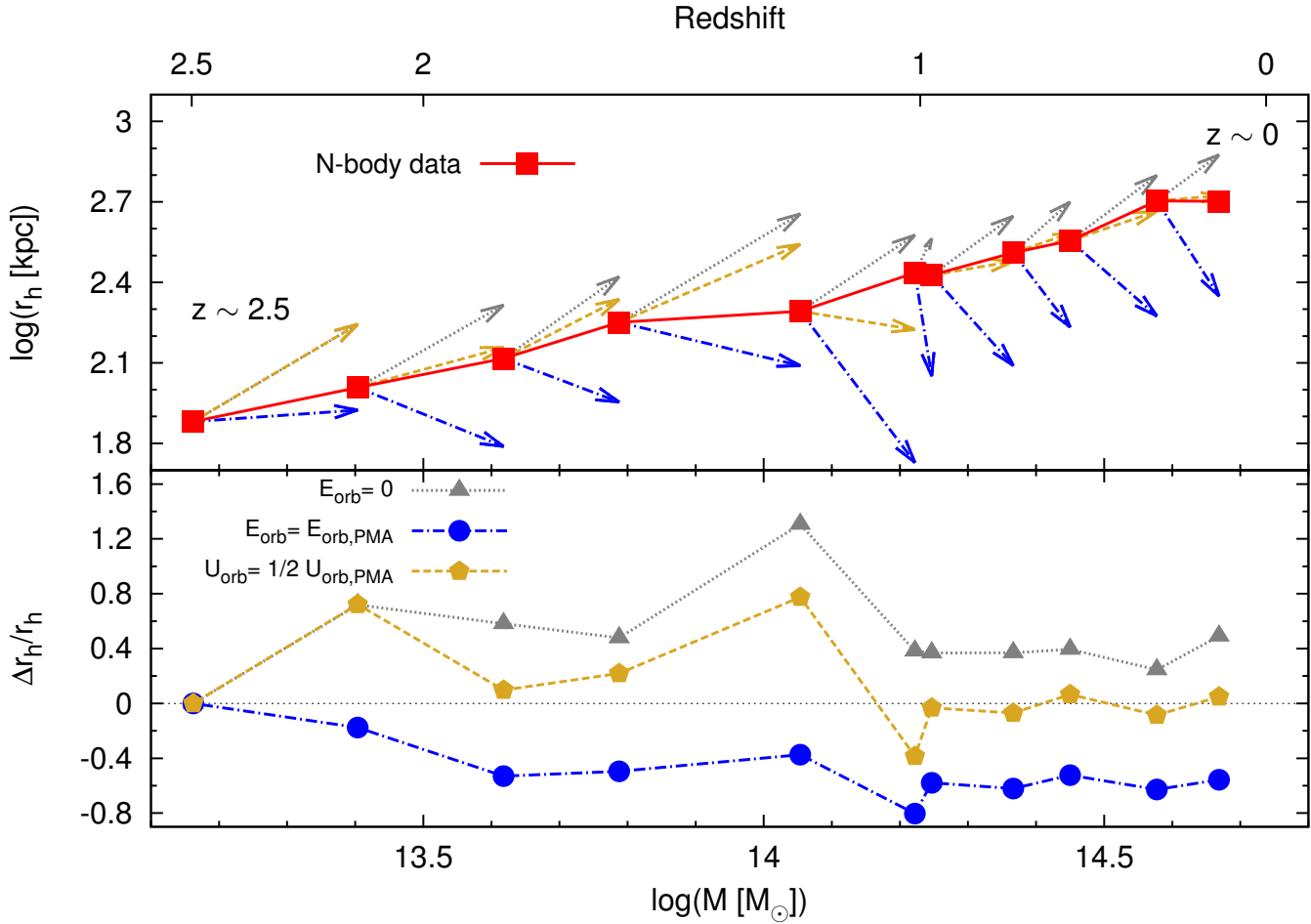}
\caption{Same as Fig. \ref{fig:evotracks_Msig}, but for the evolution
  in the $M-\rh$ plane.  }
\label{fig:evotracks_Mrh}
\end{figure*}

The orbital potential energy computed in the PMA ($\UPMA$) is
always larger (in modulus) than the actual orbital potential
energy $\Uorb$, computed for the extended objects.
As an example, one can think of a toy model in which the two
particle systems are initially very far away, where the PMA is
justified, and then they start to get nearer and nearer up to the
limit in which the centres of mass of the two system coincide: in this
limit the orbital potential $\Uorb$ in the PMA diverges, while that of
the extended system is still finite.

We tried to obtain a better estimate of the orbital energy by
empirically correcting $\UPMA$ as follows.
We computed the relative distance $\drel$ of a merger as the distance
of the centres of the haloes in the snapshot before of the
merging, i.e., is the latest snapshot in which the two haloes are
distinct. Calculating the distribution of $\drel$ of the collisions in
our simulation, we find that it peaks at $\drel \simeq \rdelta$ of the
biggest merging halo.  Assuming an NFW \citep[see][]{NFW1997} profile
for the dark haloes, we made some experiments in order to evaluate the
overestimate of $|\Uorb|$ in the PMA varying the parameters of the
haloes (such as mass ratio, size ratio and concentration) and, most
important, the relative distance between their centres of mass. We
used \Gadget to calculate the $\Uorb$ of the encounter
between the two systems and we compared it to that calculated in the
PMA.  We find that, quite independently of the structural
parameters of the haloes, in a range of relative distances consistent
with that measured in our simulation, $\Uorb \simeq \UPMA/2$.

\subsubsection{Application to $N$-body data}

 In the hypothesis that mergers are the driving mechanism for halo
  evolution, it is possible, with equation \eqref{eq:sig_nonpar}, to
  predict the evolutionary track of a halo in the $M-\sigmazero$ plane
  across different snapshots in the simulation. For a given halo,
starting from $M$ and $\sigmazero$ of the most massive progenitor, we
considered one at a time every merger in the halo merger tree and we
used for each of them equation \eqref{eq:sig_nonpar} to predict the
evolution of $\sigmazero$ after the collisions.
In Fig. \ref{fig:evotracks_Msig} (top panel) we show the evolution of
a representative halo in the $\sigmazero-M$ plane and we compare it with
the predictions of three dry-merging models based on equation
(\ref{eq:sig_nonpar}): the parabolic-merger model (i.e.  $\Eorb=0$), the
PMA model ($\Eorb=\EPMA$, where $\EPMA$ is the orbital energy computed in
the PMA) and the corrected PMA model
($\Uorb = \UPMA/2$; see Section \ref{sec:orbitalpar}). The relative
errors of the models with respect to the $N$-body data (bottom panel
of Fig. \ref{fig:evotracks_Msig}) indicate that in the majority of the
cases, the parabolic merger approximation tends to underestimate the
$\sigmazero$ evolution by a factor $\approx 20-30\%$, while
$\Eorb=\EPMA$ model tends to overestimate its
growth up to $\approx 80\%$, probably due to the overestimate
of $|\Uorb|$ introduced by the PMA. When we apply the empirical
correction we find a much better agreement with the $N$-body data.
We have checked that the resulting behaviour is fairly independent
of the halo considered (i.e., on the halo mass at $z\sim 0$).

Using equation \eqref{eq:rh_nonpar}, where we compute $\sigmavf$ as in
equation \eqref{eq:sig_nonpar}, we are able to predict also the
evolution in size of the halo.  The results are summarized in
Fig. \ref{fig:evotracks_Mrh}. The
parabolic merging model always tends to overestimate the actual size
growth of the halo: such trend is in agreement with the
analytical expectation of strong growth of the halo size (see
Section \ref{sec:merg_analytic}).  On the other hand, the $\Eorb = \EPMA
$ model is underestimating the $\rh$ evolution.
Applying the empirical correction to the orbital energy in the PMA,
we find a much better representation of the $N$-body data with respect
to previous cases (see the bottom panel of Fig. \ref{fig:evotracks_Mrh}),
in agreement with our previous results on the velocity dispersion
evolution (see Fig. \ref{fig:evotracks_Msig}). 

The results above deserve a further comment: since in the code we are
using energy is conserved \citep[see][]{Springel2005}, in principle,
if we took into account all the possible complications to the merging
model we would reproduce the actual measured
evolution. Here we focus on the effect of one of these
possible complications, namely the orbital energy. Other authors have
used this approach before, studying for example the effect of escapers
\citep[see e.g., \citetalias{Nipoti+2003};][]{Hilz+2012}: in particular, \cite{Hilz+2012}
found that mass loss can be very important in reproducing the size growth
of collisionless systems in binary mergers simulations.
Our effort is complementary to such works: we added a detailed study of
the importance of the orbital energy, finding that taking into account
such effect in a simple dry merging model can be \emph{crucial} in order to
reproduce the halo evolution in a cosmological context.

\section{Implications for the size evolution of early-type galaxies}
\label{sec:DMdriven_evo}

In the previous sections we have described the evolution in size and
velocity dispersion of the population of DM haloes. This evolution
is qualitatively similar to that of the observed population of ETGs.
The aim of this section is to compare quantitatively our $N$-body data
to observations. Since there are no baryons in the simulation, we
need to populate our DM haloes with galaxies, by assigning the
stellar mass $\Mstar$ and the stellar effective radius $\re$, which
can be done using currently available prescriptions for the
stellar-to-halo mass relation (SHMR) and the stellar-to-halo size
relation (SHSR).

\subsection{The stellar-to-halo mass relation (SHMR)}
\label{sec:mh-mstar}

  A critical point in this work is the assignment of stellar masses
  to the dark haloes of our $N$-body simulation. To do so, we need to
  assume a SHMR, i.e., a function that associates a stellar mass $\Mstar$
  to each given halo mass $M$ at a given redshift $z$. Many prescriptions
  are available at the time of this writing for this function \citep[see
    e.g.,][]{Behroozi+2010,Wake+2011,Leauthaud+2012,Moster+2013}, but
  the detailed properties of the SHMR are still uncertain and debated.

\begin{figure}
\includegraphics[width=\hsize,height=70mm]{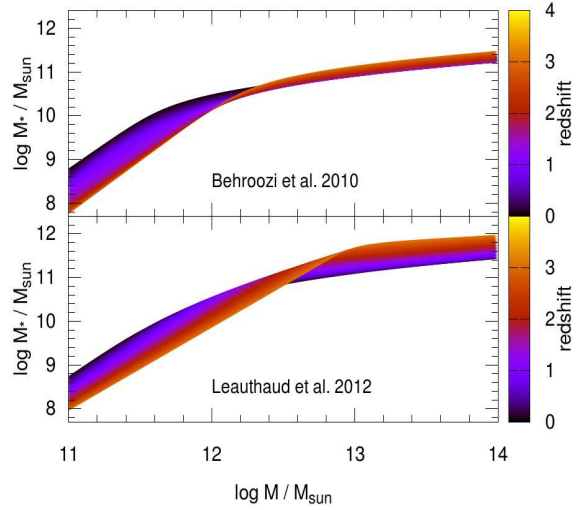}
\caption{Stellar mass $\Mstar$ as a function of the halo mass $M$ and
  redshift according to the prescriptions of
  \citetalias{Behroozi+2010} (top panel; here used in Model 1) and
  \citetalias{Leauthaud+2012} (bottom panel; here used in Model 2).  }
\label{fig:Ms-Mh_models}
\end{figure}

In order to account for the uncertainties in the SHMR, we use two
different models of SHMR: Model $1$, adopting the prescription of
\citeauthor{Behroozi+2010} \citepalias[\citeyear{Behroozi+2010},
  hereafter][] {Behroozi+2010} and Model $2$, adopting the
prescription of \citet[][hereafter
  \citetalias{Leauthaud+2012}]{Leauthaud+2012}.
A graphical representation of such models can be found
in Fig. \ref{fig:Ms-Mh_models}: the SHMR is plotted in different
colours in the redshift range $0 \leq z \leq 4$. The functional
  forms and parameters of the two prescriptions used here are
  summarized in section~3.2.1 of \citetalias{Nipoti+2012}.  Throughout the
  paper we adopt a \cite{Cha03} initial mass function. As in
\citetalias{Nipoti+2012}, for simplicity, we do not take into account the
scatter of the SHMR: to each halo we assign an $\Mstar$ which is the
mean value of the distribution.

We consider a subset of the DM halo population presented in Section \ref{sec:definitions}.
We cut our sample of objects to have a stellar mass $\log\Mstar/\Msol
\geq 10.5$ and such that no halo exceeds $M \simeq 4\times 10^{13}\Msol$.
The stellar mass lower limit is so that the population is dominated by
ETGs, since at high stellar mass the fraction of ETGs over the total
number of galaxies is larger. The halo mass upper limit is
motivated by the fact that DM haloes with mass larger than $M =
4\times 10^{13}\Msol$ are more likely to host groups of galaxies
than single ETGs.
We notice here that even if the two models have the same upper limit
in halo mass, they have a different upper limit in stellar mass,
because of the different model of SHMR used: Model $2$ has a steeper
slope and a higher normalization at the high-mass end, resulting in
more massive galaxies associated to the same halo mass with respect to
Model $1$.  The stellar mass ranges used here are: $3.2 \times
10^{10}\Msol \leq \Mstar \leq 1.4 \times 10^{11}\Msol$ (corresponding
to $8.5 \times 10^{11}\Msol \leq M \leq 4 \times 10^{14}\Msol$ in halo
mass) for Model $1$ and $3.2 \times 10^{10}\Msol \leq \Mstar \leq 2
\times 10^{11}\Msol$ (corresponding to $9 \times 10^{11}\Msol \leq M
\leq 4 \times 10^{14}\Msol$ in halo mass) for Model $2$.

\subsection{The stellar-to-halo size relation (SHSR)}
\label{sec:rh-re}
After having characterized every DM halo with a stellar mass under
the assumption of a SHMR, the second step is to assign a size, namely an
effective radius $\re$, to the stellar component. Assuming a
reasonable form for the SHSR is not trivial and at this time there is
not yet a prescription that can be taken as a reference.

However, recently \citet[][following the theoretical work
of \citetalias{Mo+1998} and \citeauthor{FallEfstathiou1980}
\citeyear{FallEfstathiou1980}]{Kravtsov2013} argued that such relation can be
measured over a wide range of stellar masses, using abundance matching
techniques to derive a functional form. \cite{Kravtsov2013} finds that
the relation between the virial radius $\rdelta$ of the host halo and
the effective radius $\re$ of the galaxy is linear.  Such result
confirms the theoretical predictions \citepalias[see][]{Mo+1998} that the
virial radius of the halo is linearly proportional to the size of the
galactic disk and that the
constant of proportionality depends of the spin parameter $\lambda = J
\, |E|^{1/2} G^{-1} M^{-5/2}$, where $J$ is the norm of the angular
momentum of the halo, $E$ is the total energy of the halo and $G$ is
the gravitational constant.  Moreover, the fact that both early-type
and late-type galaxies follow this linear SHSR with a scatter of $\sim
20\%$ \citep[see][]{Kravtsov2013} can be interpreted as
the fact that the \citetalias{Mo+1998} model not only works for disk
galaxies, but it represents a general behaviour of all types of massive
galaxies.

\begin{figure}
\includegraphics[width=\hsize]{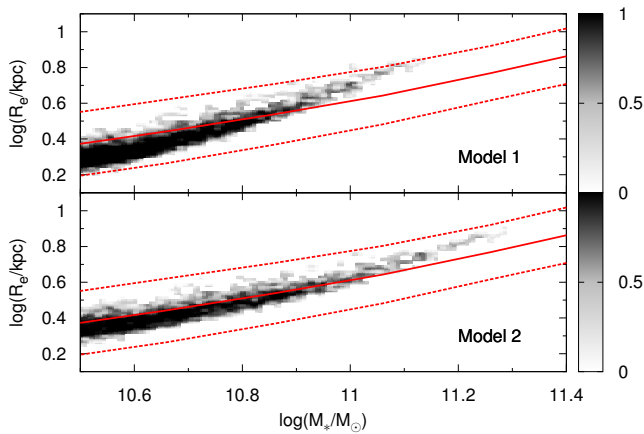}
\caption{Distribution in the plane stellar mass-effective radius of
  the sample of model galaxies obtained by populating the $z \simeq 0$
  simulated DM haloes according to Model $1$ (top panel) and Model $2$
  (bottom panel). In both panels, the gray scale is proportional to
  the logarithm on the number counts of haloes in the binned plane and
  the solid line is the best-fit relation of \citet{Shen+2003}, for
  SDSS galaxies, with the corresponding $1\sigma$ scatter (dashed
  lines).  }
\label{fig:Ms-Re_z=0}
\end{figure}

Here we assume $\re\propto\rh$, where $\rh$ is the halo half-mass
radius, defined in Section \ref{sec:definitions}.  However, in Section
\ref{sec:behaviour_diffdef} we showed that $\rh\propto\rdelta$, on
average, so our assumption is consistent with the results of
\cite{Kravtsov2013}.  As a check to this hypothesis, we compare
the stellar mass-effective radius correlation for our model galaxies
(i.e., $\Mstar-\re$) at $z=0$ with that observed in the local Universe,
taking as reference the $\Mstar-\re$ correlation of the ETGs in the
Sloan Digital Sky Survey \citep[SDSS;][]{Shen+2003}. In
Fig. \ref{fig:Ms-Re_z=0} we plot the distribution of the effective
radius as a function of the stellar mass for our model galaxies in
Models $1$ and $2$ (see Section \ref{sec:mh-mstar}), and we compare
it to the best-fit relation of \cite{Shen+2003}. In
particular, we have used $\re/\rh = 0.031$ for Model $1$ and $\re/\rh
= 0.042$ for Model $2$, which are in reasonable agreement with the
expectations of \citetalias{Mo+1998} and the results of \cite{Kravtsov2013},
given that on average we find $\rh/\rdelta \simeq 0.82$ in the
simulated haloes. Overall, Fig.~\ref{fig:Ms-Re_z=0} shows that
the distribution of the $z\simeq 0$ model galaxies in the
$\re-\Mstar$ plane is consistent with that of SDSS galaxies. In more
detail, Model $2$, with a best-fit $\re\propto\Mstar^{0.64}$,
appears to reproduce better the SDSS data (best-fit
$\re\propto\Mstar^{0.56}$) than Model $1$ (best-fit
$\re\propto\Mstar^{0.84}$).

When comparing our models to observations (see Section
\ref{sec:sizeevo}) we will assume that the ratio $\re/\rh$ is
independent of $M$ and $z$. We will study only the redshift evolution
of size ratios [namely, $\re(z)$ normalized to $\re(z \simeq 0)$], so
the results are independent of the actual value of $\re/\rh$.
We note that an underlying assumption of our models is that baryons
do not affect substantially the structural evolution of DM haloes.
Though the stellar and DM components are expected to affect each
other significantly, even in dissipationless mergers \citepalias[see e.g.,]
[]{Hilz+2013}. However, our results are not sensitive to this effect as
long as it is independent on halo mass.

\subsection{Size evolution of early-type galaxies: comparing models with observations}
\label{sec:sizeevo}

\begin{figure*}
\includegraphics[width=\hsize]{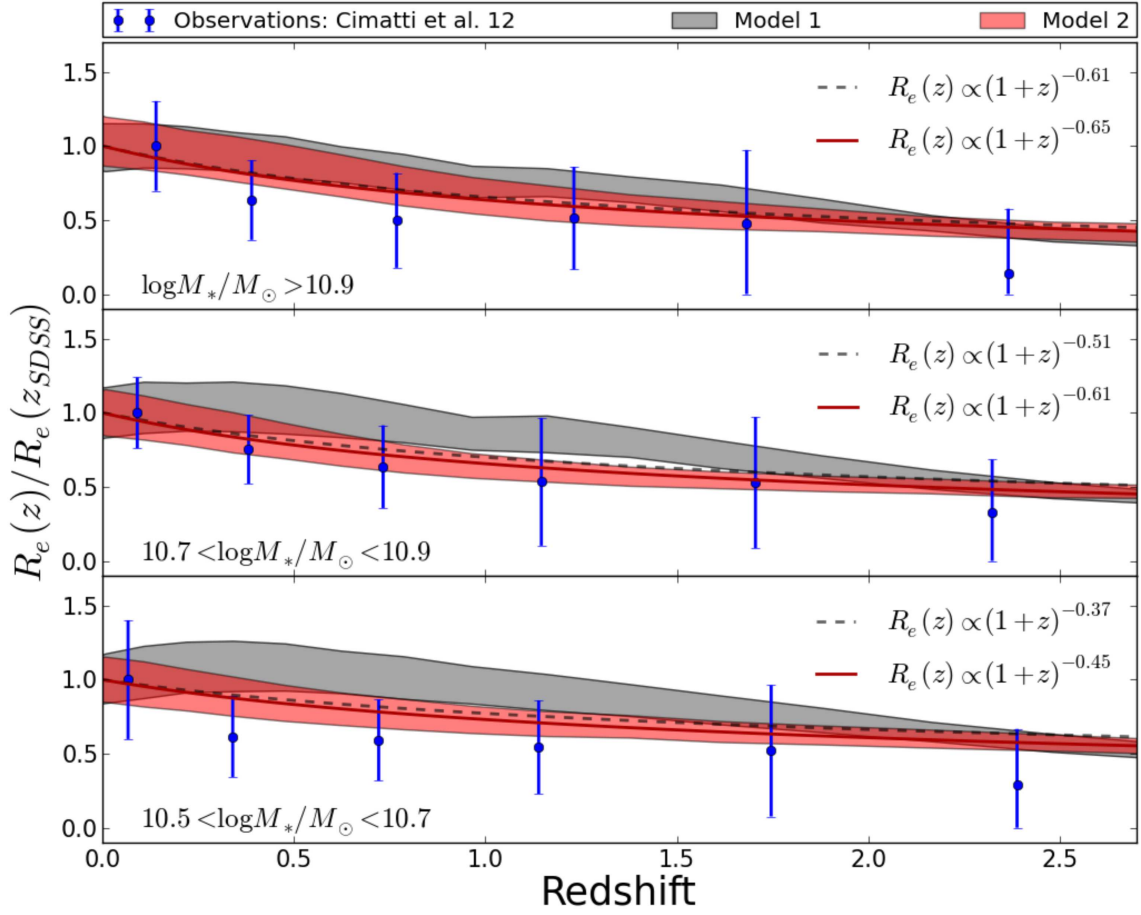}  
\caption{ Average effective radius $\re$ as a function of redshift
    for simulated galaxies of Model 1 (gray bands) and Model 2 (red
    bands) of the present work and for observed ETGs of
    \citetalias{Cimatti+2012} (blue filled circles). Overplotted
    are also the best-fit power-laws $\re\propto(1+z)^{\gamma}$
    to Model $1$ (gray dashed line) and Model $2$ (red solid line).
    Each panel refers to the indicated stellar mass interval and the
    radii are normalized to the average $\re$ of SDSS galaxies in that
    mass interval. The models are anchored to the lowest-$z$ (i.e. SDSS)
    observational points (see text). The vertical bars and the widths
    of the bands indicate one standard deviation.}
\label{fig:size-evo}
\end{figure*}

Here we compare the size evolution of our model galaxy sample
(built from the $N$-body data as described in
Sections~\ref{sec:mh-mstar} and \ref{sec:rh-re}) with that of the
observed population of ETGs. In particular, we take as reference
observational sample the collection of ETGs in the redshift range $0
\lesssim z \lesssim 3$ presented by \citeauthor{Cimatti+2012}
(\citeyear{Cimatti+2012}, hereafter \citetalias{Cimatti+2012}).
We show such comparison in Fig. \ref{fig:size-evo}: we compute the average
size of the model galaxy sample at different times ($16$ snapshots of
the $N$-body simulation in the range $0 \leq z \leq 3$) and we
normalize it to that at the mean redshift of the SDSS. Following
\citetalias{Cimatti+2012}, we plot the evolution in three different
mass bins.  In Fig. \ref{fig:size-evo} we are showing a sort of
\emph{backward} evolution: we are normalizing our models to be in
agreement with the data at $z \simeq 0$ and we follow the evolution of
the average size of the model galaxies at higher redshift. Our choice
is motivated by the fact that we anchor the models to observations in the
local Universe, which are more numerous with respect to $z \simeq 2.5-3$ data.

The distribution of our model galaxies significantly
overlaps with that of the observed ETGs of \citetalias{Cimatti+2012},
but the models suffer from a systematic underestimate
of the size growth in all the mass bins, in particular at $z>2$. At
$z<2$ the models \citepalias[especially Model $2$, which is based
on][]{Leauthaud+2012} are consistent within the scatter with
the observational data, suggesting that the host haloes
actually left their footprint in the stellar density at the time of galaxy
quenching and that the after-quenching evolution of the galaxies mimicked
that of the haloes.
To give reference numbers, for both models, we
computed the ratio $\re(z \simeq 2.5)/\re(z \simeq 0)$ in the three
different mass bins finding that from $z \simeq 2.5$ to $z \simeq 0$ the
average effective radius increases of roughly a factor of $\approx 2$ in
the lowest mass bin ($\log \Mstar/\Msol < 10.7$) and a factor of $\approx 3$
in the highest one ($\log \Mstar/\Msol > 10.9$).
For comparison, most observational data agree on an
evolution of the average effective radius of the ETG population of a
factor $2-5$ from $z \gtrsim 2$ to the present \citepalias[see e.g.,][]
{Cimatti+2012}. 

In Table \ref{tab:models} we report the best-fit values of $\gamma$,
where $\re\propto(1+z)^{\gamma}$, for both models in the range $0 \lesssim z
\lesssim 2.5$ and in the three mass bins.
We find values of $\gamma$ similar to those in \citetalias{Cimatti+2012}
when they exclude the $z>2$ data, while the models fail at reproducing the
observed slopes which take into account also the higher redshift points.

Our analysis shows that, provided that the galaxy formation process
produces a linear SHSR, the observed size evolution of the population
of ETGs up to $z\simeq 2$ could be explained by the underlying size
evolution of the halo population, in the sense that quiescent galaxies
mimic the host halo evolution.
At $z > 2$ the average size of the observed population of ETGs
evolves significantly faster than predicted by our simple models
\citepalias[in agreement with \citetalias{Cimatti+2012};][]{Nipoti+2012}.
This difference at $z\gtrsim 2$ can possibly give us some insights
into the role dissipative effects, such as star formation or active galactic
nuclei (AGN) feedback: it might not be a mere coincidence that at
$z \approx 2$ there are the peaks of the cosmic star formation rate
\citep{Madau+1996,Lilly+1996} and of AGN activity \citep[see e.g.,][]
{MerloniHeinz2008,Gruppioni+2011}.

Overall, our results are in agreement with those of previous
investigations \citep[\citetalias{Cimatti+2012,Nipoti+2012};][]{Newman+2012},
which are in a sense complementary to the present work. For instance, we
note that the approach used here is different from that of
\citetalias{Cimatti+2012}, who (following \citetalias{Nipoti+2012}) treat in
their model only the evolution of individual galaxies and do not
include the contribution of galaxies that have become quiescent at
relatively low redshift. In other words, when comparing models to
observations, \citetalias{Cimatti+2012} do not try to account for the
so-called progenitor bias \citep[see e.g.,][]{Saglia+2010,Carollo+2013}
because they assume that the observed population of high-$z$ ETGs is
representative of the progenitors of present-day ETGs. Our approach, though
simple, should be more robust against the progenitor bias: in our sample of
simulated objects, at any redshift, we can have in principle both
galaxies that have just stopped forming stars (and that are
identified for the first time as ETGs) and galaxies that became
quiescent much earlier. However, it must be stressed that our model
is limited by the fact that we are assuming that galaxies grow in size
and mass so that the SHMR and SHSR are reproduced at all redshifts, without
having specified any underlying physical model for such growth. In
this respect, a more physically justified approach is that of
\citetalias{Cimatti+2012} and \citetalias{Nipoti+2012}, who assume that the size
and mass growth of ETGs is driven by dry mergers, finding that, under
this hypothesis, the SHMR inferred from observations is not
necessarily reproduced (see \citetalias{Nipoti+2012}). In summary, our
results, combined with those of similar previous works, suggest a
scenario in which, at least up to $z \simeq 2$, the observed growth of
ETGs reflects the underlying growth of their host DM haloes.

\subsection{Velocity dispersion evolution of early-type galaxies: comparing models with observations}
\label{sec:sigmaevo}

In this Section we compare the stellar velocity dispersion evolution of
our model ETGs with that observed up to $z \simeq 2$. For such comparison
we rely on the data collection of ETGs from \cite{Belli+2013} and
\cite{vandeSande+2013}.
We use a very simple recipe to get the stellar velocity dispersion
of the model galaxies: consistent with our choice for the assignment
of $\re$ (i.e., $\re\propto\rh$), we assume a scaling with the halo properties
of the form $\sigmastar^2 \propto \Mstar(M)/\re(\rh)$, where we use the
SHMR (Section \ref{sec:mh-mstar}) and SHSR (Section \ref{sec:rh-re})
from Model $1$ and $2$.

\begin{figure}
\includegraphics[width=\hsize]{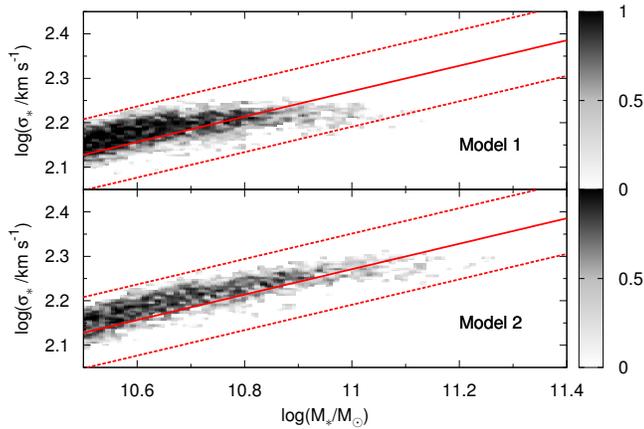}
\caption{Same as Fig. \ref{fig:Ms-Re_z=0}, but in the stellar mass-velocity
  dispersion plane. In both panels,
  the solid line is the best-fit relation of \citet{Hyde&Bernardi2009}, for
  SDSS galaxies, with the corresponding $1\sigma$ scatter (dashed
  lines).}
\label{fig:Ms-sigs_z=0}
\end{figure}

We check this recipe for ETGs at $z \simeq 0$: in Fig.
\ref{fig:Ms-sigs_z=0} it is plotted the stellar mass-velocity
dispersion relation at $z\simeq 0$ for our model ETGs which we
compare to SDSS observations data from \cite{Hyde&Bernardi2009}.
Overall, both models are able to represent fairly well the SDSS data
within their uncertainties. However, we notice that Model $2$ works
systematically better than Model $1$ in reproducing the observed
$\Mstar-\sigmastar$: Model $1$ is best-fitted by $\sigmastar\propto
\Mstar^{0.11}$, while for Model $2$ we find $\sigmastar\propto\Mstar^{0.2}$,
closer to the \cite{Hyde&Bernardi2009} relation
$\sigmastar\propto\Mstar^{0.286}$.

\begin{figure}
\includegraphics[width=\hsize]{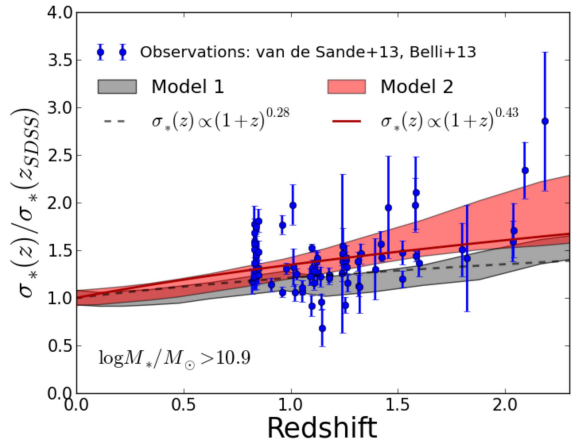}  
\caption{Average velocity dispersion $\sigmastar$ as a function of redshift
    for simulated galaxies of Model $1$ (gray bands) and Model $2$ (red
    bands) of the present work and for observed ETGs of
    \citet{vandeSande+2013} and \citetalias{Belli+2013} (blue filled circles).
    Here we restrict to stellar masses $\log\Mstar / M_\odot > 10.9$.
    Overplotted are also the best-fit power-laws $\sigmastar\propto(1+z)^{\delta}$
    (gray dashed line and red solid line, respectively to Model $1$ and Model $2$).
    The models are anchored to the lowest-$z$ (i.e. SDSS)
    observational points (see text), as in Fig. \ref{fig:size-evo}.
    The vertical bars and the widths of the bands indicate one standard deviation.}
\label{fig:sigma-evo}
\end{figure}

Figure \ref{fig:sigma-evo} shows the average velocity dispersion (\emph{backward})
evolution of our model galaxies in the redshift range $0 \lesssim z \lesssim 2$.
We compute the average $\sigmastar$ as a function of
redshift, we anchor the data from \citetalias{Belli+2013} and \cite{vandeSande+2013}
to a reference value for the local Universe  taken from \cite{Hyde&Bernardi2009}
and we directly compare the backward evolution with data available for individual ETGs.
Qualitatively we find a reasonable agreement with observations of high-mass
(i.e., $\log\Mstar/\Msol>10.9$) passive galaxies up to $z\simeq 2$.
The predictions, expecially those of Model $2$, are able
to reproduce the evolutionary trend in the observations, even though there
is significant scatter. Quantitavely, we find that Model $2$ evolution
is well fitted by $\sigmastar\propto(1+z)^{\delta}$, with $\delta =
0.43 \pm 0.07$, while for Model $1$ we find $\delta = 0.28
\pm 0.05$. For comparison, fitting the individual observations
with a similar power-law evolution gives us $\delta = 0.36 \pm 0.02$,
which is consistent within the uncertainties at least with Model $2$.
The model galaxies have decreased on average their velocity dispersion by a
factor $\sigmastar (z\simeq 2)/\sigmastar (z\simeq 0) \approx 1.4-1.8$,
for comparison \cite{vandeSande+2013} quote an evolution in their sample
of data of a factor $\approx 1.4-1.7$.

In Table \ref{tab:models} we summarize the results of fitting the velocity
dispersion evolution in both models with simple power-laws, in the range
$0 \lesssim z \lesssim 2.5$ and in the three stellar mass bins considered
in Section \ref{sec:DMdriven_evo}. The numbers reported in Table
\ref{tab:models} and the trends shown by Fig. \ref{fig:size-evo} and
\ref{fig:sigma-evo} are also in good agreement with previous results of the
\emph{average} size and velocity dispersion evolution of stellar systems
in hydrodynamical cosmological simulations \citep[see e.g.,][]{Oser+2012}.

The above results on the velocity-dispersion evolution depend on the assumed
assignement of $\sigmastar$ to haloes (i.e., $\sigmastar^2\propto\Mstar/\re$).
A different choice could be $\sigmastar\propto\sigmazero$ (where $\sigmazero$ is
the halo velocity dispersion); however, we have verified that assuming a scaling
of that type leads to a very poor comparison of the $\Mstar-\sigmastar$ relation
with data for the local Universe. 
This suggests that the evolution of the velocity dispersion of the stellar component
is somewhat decoupled from that of the dark component: while $\sigmazero$ increases
with time for individual haloes in our sample (see Fig. \ref{fig:wholevo_Msig_Mrh}),
viceversa $\sigmastar$ of individual model galaxies tends to decrease as a function
of time, if one assumes $\sigmastar^2\propto\Mstar/\re$.

The same trend of velocity dispersion decreasing with time for individual galaxies
is found in the cosmological hydrodynamical simulations of \citet[][Oser private
communication]{Oser+2012}. However, it must be noted that \cite{Dubois+2013}, 
who also studied the evolution of early-type galaxies using cosmological
hydrodynamical simulations, found that the velocity dispersion of individual
galaxies increases with time.

\begin{table*}
\begin{center}
\caption{Summary of the best fitting power-laws $\re(z) \propto (1+z)^{\gamma}$ and
    $\sigmastar(z)\propto (1+z)^{\delta}$ to our models in the range
    $0 \lesssim z \lesssim 2.5$, for different stellar mass bins (see Section \ref{sec:sizeevo}
    and \ref{sec:sigmaevo}).}

\begin{tabular}{lccccc}
\label{tab:models}
 & & \multicolumn{2}{c}{Model $1$} & \multicolumn{2}{c}{Model $2$} \\
 \hline \hline \noalign{\vspace{0.3mm}}
  Stellar mass & & $\gamma$ & $\delta$ & $\gamma$ & $\delta$ \\
 \hline \noalign{\vspace{0.3mm}}
  $\log \Mstar/\Msol>10.9 $ & & $-0.61 \pm 0.07$ & $0.28 \pm 0.05$ & $-0.65 \pm 0.09$ & $0.43 \pm 0.07$ \\ \noalign{\vspace{1.5mm}}
  $10.7 < \log \Mstar/\Msol<10.9 $ & & $-0.51 \pm 0.06$ & $0.18 \pm 0.05$ & $-0.61 \pm 0.06$ & $0.20 \pm 0.06$ \\ \noalign{\vspace{1.5mm}}
  $10.5 < \log \Mstar/\Msol<10.7 $ & & $-0.37 \pm 0.06$ & $0.04 \pm 0.05$ & $-0.45 \pm 0.06$ & $0.01 \pm 0.05$ \\ \noalign{\vspace{0.3mm}}
 \hline
\end{tabular}
\end{center}

\end{table*}

\section{Summary and Conclusions}
\label{sec:concl}

Motivated by the observational finding that ETGs are, on
average, more compact at higher redshift, we have explored the
hypothesis that such evolution is mainly driven by the
systematic redshift-dependence of the structural properties
of their host DM haloes. Using a cosmological $N$-body simulation,
we have followed the evolution of the structural and
kinematical properties of a DM halo population in the $\Lambda$-CDM
framework, focusing on the halo mass range $10^{11} \lesssim M/\Msol
\lesssim 5 \times 10^{14}$. Starting from a sample of simulated
haloes, we have built a sample of model ETGs and we have compared
the redshift evolution of their sizes with that of observed
galaxies.

The main results can be summarized as follows:
\begin{itemize}
 \item At $z=0$ the haloes are well represented by $\sigmazero\propto
 M^{0.329 \pm 0.001}$ and $\rh\propto M^{0.320 \pm 0.002}$ at $z=0$,
 where $\rh$ is the half-mass radius and $\sigmazero$ is the central
 velocity dispersion. These global correlations are remarkably similar
 to those predicted for the virial quantities of the haloes (namely,
 $M\propto\sigmav^3$ and $M\propto\rdelta^3$), meaning that there is
 not significant non-homology in the halo population.
\item The slopes of the $M-\sigmazero$ and $M-\rh$ correlations depend
  only slightly on $z$, but their normalizations evolve significantly
  with $z$ in the sense that, at fixed mass, higher-$z$ haloes have
  smaller $\rh$ and higher $\sigmazero$. For instance, at fixed
  $M=10^{12}\Msol$ we find $\sigmazero\propto(1+z)^{0.35}$ and
  $\rh\propto(1+z)^{-0.71}$.
\item The redshift evolution of the halo scaling laws is driven
  by individual haloes growing in mass following evolutionary tracks
  $\sigmazero\propto M^{0.2}$ and $\rh\propto M^{0.6}$.
  So, while individual haloes grow in mass, their
  velocity dispersions increase slowly and their sizes grow rapidly. 
\item The size and velocity dispersion evolution of individual haloes
   is successfully described by simple dissipationless merging models,
   in which a key ingredient is the (typically negative) orbital
   energy of the encounters.
\item We compare our $N$-body data with observations of ETGs in the
   redshift range $0\lesssim z\lesssim 3$, by populating the DM haloes
   with galaxies assigning to each halo a stellar mass, an
   effective radius and a stellar velocity dispersion.
   We find that the size and velocity dispersion evolution of our model
   galaxies is in reasonable agreement with the evolution observed for
   ETGs at least up to $z \simeq 2$. At $z>2$ the observed size growth
   is stronger than predicted by our simple models.
\end{itemize}

  The above findings suggest a scenario in which the size and velocity
  dispersion scaling laws of ETGs derive from underlying
  scaling laws of the DM haloes. 
  Overall, the results of the present work give further support to the 
  idea of a halo-driven evolution of ETGs: galaxy structural and dynamical
  properties are related to that of their haloes at the time of quenching
  and the further ETG evolution mimics that of haloes.
  Of course, the present approach is limited by the fact that we do not
  include a self-consistent treatment of baryonic physics. In a
  forthcoming study, we plan to investigate this problem by adding the
  baryonic evolution in the simulations.

\section*{Acknowledgements}

We are grateful to Andrea Cimatti for useful discussions and
to Michele Trenti for providing his modified version of
the initial conditions generator code {\sc GRAFIC2}. LP is also
grateful to the \AHF
community for helpful discussions about the halo finder. Part of the
numerical calculations were run on the UDF Linux Cluster at the Space
Telescope Science Institute and part on the IBM Cluster SP$6$ at
CINECA (Bologna). We acknowledge the CINECA Awards N. HP10C2TBYB
(2011) and N. HP10CQFATD (2011) for the availability of high
performance computing resources and support.  MS and LP were partially
supported by the JWST IDS grant NAG5-12458.  LC and CN acknowledge
financial support from PRIN MIUR 2010-2011, project \textquotedblleft
The Chemical and Dynamical Evolution of the Milky Way and Local Group
Galaxies\textquotedblright, prot. 2010LY5N2T.  All the numerical
computations and the graphic representations have been produced using
exclusively open source software.

\end{document}